\documentclass{aa}

\usepackage{graphicx}
\usepackage{txfonts}
\usepackage{lipsum}
\usepackage{xcolor}
\usepackage{subcaption}
\usepackage{hyperref}
\usepackage{amsmath}
\usepackage{tabularx, multirow}
\usepackage{soul}

\def\kmsmpc{km s$^{-1}$ Mpc$^{-1}$}

\def\ergs{erg s$^{-1}$}

\def\Hzerg{Hz erg$^{-1}$}

\def\msun{\ifmmode M_{\odot} \else M$_{\odot}$\fi}
\def\msunyr{\ifmmode M_{\odot} {\rm yr}^{-1} \else M$_{\odot}$ yr$^{-1}$\fi}
\def\msunyrvol{\ifmmode \msunyr {\rm Mpc}^{-3} \else \msunyr Mpc$^{-3}$\fi}
\def\zsun{\ifmmode Z_{\odot} \else Z$_{\odot}$\fi}
\def\lsun{\ifmmode L_{\odot} \else L$_{\odot}$\fi}

\newcommand{\oh}{\ifmmode 12 + \log({\rm O/H}) \else$12 + \log({\rm
O/H})$\fi}

\newcommand{\oiii}{\ifmmode \text{[O~{\sc iii}]} \else[O~{\sc iii}]\fi}

\newcommand{\ha}{\ifmmode \text{H}\alpha \else H$\alpha$\fi}
\newcommand{\hb}{\ifmmode \text{H}\beta \else H$\beta$\fi}
\newcommand{\oiiihb}{\text{\oiii+\hb}}
\newcommand{\oiiifull}{\Oiii}
\newcommand{\oiiihbfull}{\Oiiihb}

\def\Oiii{\ifmmode \oiii\lambda\lambda4960,5008 \else\oiii$\lambda\lambda4960,5008$\fi}
\def\Oiiib{\ifmmode \oiii\lambda5008 \else\oiii$\lambda5008$\fi}
\def\Oiiit{\ifmmode \oiii\lambda4960 \else\oiii$\lambda4960$\fi}
\def\Oiiihb{\hb+\Oiii}

\newcommand{\mstar}{\ifmmode M_\star \else $M_\star$\fi}
\newcommand{\muv}{\ifmmode M_\text{UV}\else $M_\text{UV}$\fi}
\newcommand{\auv}{\ifmmode A_{\rm UV} \else $A_{\rm UV}$\fi}
\newcommand{\luv}{\ifmmode L_{\rm UV} \else $L_{\rm UV}$\fi}
\newcommand{\lir}{\ifmmode L_{\rm IR} \else $L_{\rm IR}$\fi}
\newcommand{\lbol}{\ifmmode L_{\rm bol} \else $L_{\rm bol}$\fi}
\newcommand{\liruv}{\ifmmode L_{\rm IR+UV} \else $L_{\rm IR+UV}$\fi}
\newcommand{\liroveruv}{\ifmmode L_{\rm IR}/L_{\rm UV} \else $L_{\rm IR}/L_{\rm UV}$\fi}
\newcommand{\nlyc}{\ifmmode N_{\rm Lyc} \else $N_{\rm Lyc} $\fi}
\newcommand{\rholyc}{\ifmmode \rho_{\rm Lyc} \else $\rho_{\rm Lyc} $\fi}
\newcommand{\chion}{\ifmmode \xi_{\rm ion} \else $\xi_{\rm ion}$\fi}
\newcommand{\chioncorr}{\ifmmode \xi_{\rm ion}^0 \else $\xi_{\rm ion}^0$\fi}
\newcommand{\Rthree}{\ifmmode R3 \else $R3$ \fi}
\newcommand{\Rthreefunc}{\ifmmode \Rthree(\muv) \else $\Rthree(\muv)$ \fi}

\newcommand{\mean}[1]{\langle#1\rangle}
\newcommand{\fesc}{\ifmmode f_\textrm{esc} \else $f_\textrm{esc}$ \fi}
\newcommand{\Rt}{\Rthree}
\newcommand{\nion}{\ifmmode\dot{N}_{\rm ion}\else$\dot{N}_{\rm ion}$\fi}
\newcommand{\xion}{\ifmmode\xi_{\rm ion}\else$\xi_{\rm ion}$\fi}

\begin{document} 

    \title{A GLIMPSE into the very faint end of the \oiiihbfull~luminosity function at $z \sim 7 - 9$ behind Abell S1063}
    \titlerunning{A GLIMPSE into the very faint end of the \oiiihb~luminosity function at $z \sim 7 - 9$}   

   \author{Damien Korber\inst{1}\thanks{Corresponding author: Damien Korber:\\ \href{mailto:damien.korber@protonmail.ch}{damien.korber@protonmail.ch}}
          \and Iryna Chemerynska\inst{3}
          \and Lukas J. Furtak\inst{4}
          \and Hakim Atek\inst{3}
          \and Ryan Endsley\inst{5}
          \and Daniel Schaerer\inst{1,2}
          \and John Chisholm\inst{5}
          \and Vasily Kokorev\inst{5}
          \and Alberto Saldana-Lopez\inst{6}
          \and Angela Adamo\inst{6}
          \and Julian B. Mu\~noz\inst{5}
          \and Pascal A. Oesch\inst{1,9}
          \and Romain Meyer\inst{1}
          \and Rui Marques-Chaves\inst{1}
          \and Seiji Fujimoto\inst{7,8}
          }
    
   \institute{Observatoire de Genève, Université de Genève, Chemin Pegasi 51, 1290 Versoix, Switzerland
        \and CNRS, IRAP, 14 Avenue E. Belin, 31400 Toulouse, France
        \and Institut d'Astrophysique de Paris, CNRS, Sorbonne Université, 98bis Boulevard Arago, 75014, Paris, France
        \and Department of Physics, Ben-Gurion University of the Negev, P.O. Box 653, Be'er-Sheva 84105, Israel
        \and Department of Astronomy, The University of Texas at Austin, Austin, TX 78712, USA
        \and  Department of Astronomy, Oskar Klein Centre, Stockholm University, AlbaNova University Center, SE-106 91 Stockholm, Sweden
        \and David A. Dunlap Department of Astronomy and Astrophysics, University of Toronto, 50 St. George Street, Toronto, Ontario, M5S 3H4, Canada
        \and Dunlap Institute for Astronomy and Astrophysics, 50 St. George Street, Toronto, Ontario, M5S 3H4, Canada
        \and Cosmic Dawn Center (DAWN), Denmark. Niels Bohr Institute, University of Copenhagen, Jagtvej 128, K\o benhavn N, DK-2200, Denmark 
        }

   \date{Received 14 August 2025 / Accepted 10 February 2026}

    \abstract{We used the ultra-deep GLIMPSE JWST/NIRCam survey to constrain the faint end of the \oiiihb~luminosity function (LF) down to $10^{39}$\ergs~at $z\sim7-9$ behind the lensed Hubble Frontier Field galaxy cluster Abell S1063. We applied a spectral energy distribution fitting on a Lyman-break galaxy selected sample of 164 lensed galaxies and measured their combined \oiiihbfull~flux to build the emission line LF.
    We found a \oiiihb~LF with a faint-end slope ($\alpha=-1.78_{-0.06}^{+0.06}$ for $z=7-8$ and $\alpha=-1.55_{-0.11}^{+0.11}$ for $z=8-9$), which is flatter than the UV LF at similar redshifts ($\alpha \le -2$) and suggests a lower number density of weak \oiiihb~emitting galaxies at fixed \muv. We analysed several possible explanations: {\em i)} a decrease in the \oiiihb-to-UV ratio due to bursty star formation histories (SFHs), {\em ii)} the effect of metallicity on the \oiii-to-\hb~ratio, or {\em iii)} signs of a faint-end turnover in the UV LF.
    Under the assumption of an evolving \oiii-to-\hb~ratio, we separated the contribution of \Oiiib~and \hb~and obtained a flatter \Oiiib~LF ($\alpha=-1.66_{-0.05}^{+0.05}$ for $z=7-8$ and $\alpha=-1.45_{-0.10}^{+0.09}$ for $z=8-9$) but steeper \hb~LF ($\alpha=-1.95_{-0.08}^{+0.08}$ for $z=7-8$ and $\alpha=-1.68_{-0.14}^{+0.13}$ for $z=8-9$). The combination of a decreasing metallicity and bursty SFH can reconcile the observed differences between the UV and \oiiihb~LF.
    By converting this LF into the ionising photon-production rate $\dot{N}_{\rm ion}$, we show that galaxies with $L_{\ha}\geq 10^{39}$\ergs\ , that is, with a~star formation rate (SFR) (\ha)$\geq 5\times 10^{-3}$ \msunyr) cause $31\%-90\%$ and $46\%-156\%$ of the ionising photon budget (at $7<z<8$ and $8<z<9$), when we assume a constant escape fraction of Lyman-continuum photon ($\fesc=0.14$). The shape of the LF further shows the negligible contribution of faint galaxies to the $\dot{N}_{\rm ion}$. Additionally, we derived the cosmic star formation rate density (SFRD), finding results consistent with previous estimates. However, the sensitivity of GLIMPSE to lower SFRs reinforces the conclusion that very faint galaxies contribute very little to $\dot{N}_{\rm ion}$ and the SFRD. Our results suggests that GLIMPSE has detected the bulk of the total \oiiihb~emission from star-forming galaxies, and that galaxies below our detection limits are likely minor contributors to cosmic re-ionisation.}

   \maketitle
   \nolinenumbers

\section{Introduction}
During its first billion years, the Universe underwent significant transformations. After a gradual cooling, the Universe entered the so-called dark ages, a period in which photons decoupled from matter, rendering the Universe opaque to our telescopes \citep{ferrara_reionization_2014, adamo_first_2025}.
This epoch ended around $z\sim30$ with the so-called epoch of re-ionisation (EoR) with the formation of massive and metal-free Population III stars \citep{ferrara_positive_1998}. These newly born objects emitted enough ionising photons to ionise the surrounding gas while dominating the recombination of hydrogen \citep{ferrara_positive_1998, bromm_formation_2013}. They emitted sufficient ionising photons to counteract recombination and ionise the surrounding gas in a patchy manner. Each bubble of ionised gas expanded and leaked more ionising photons, which ultimately re-ionised the whole Universe \citep[e.g.][]{zaroubi_epoch_2013, daloisio_large_2015, bosman_hydrogen_2022, robertson_galaxy_2022, korber_pinion_2023, meyer_probing_2025}. The EoR is expected to have ended around $z\sim5-6$ according to measurements of the Gunn-Peterson effect in quasar spectra \citep[e.g.][]{becker_evidence_2001, fan_constraining_2006, keating_long_2020, bosman_hydrogen_2022} and the decreasing fraction of Lyman $\alpha$ emitting galaxies detected above $z>6$ \citep{stark_keck_2011, schenker_keck_2012}. While galaxies are often considered as primary drivers of re-ionisation, the relative importance of fainter or brighter galaxies remains debated. A first scenario gives a prime role to faint galaxies because despite their small individual contribution to the ionising photon budget, their high number density might produce enough to re-ionise the Universe \citep[e.g.][]{oesch_udf05_2009, finkelstein_conditions_2019, yeh_thesan_2023, atek_most_2024, simmonds_ionizing_2024}. Another scenario gives more weight to bright galaxies, which have a large individual contribution despite their rarity \citep{naidu_rapid_2020}. However, some studies also suggested that active galactic nuclei (AGNs) might also play a significant role in re-ionisation \citep{dayal_reionization_2020, madau_cosmic_2024, maiolino_jades_2024, grazian_what_2024, singha_faint_2025}.

Until recently, probing galaxies during the EoR was hindered by technical limitations. The \textit{Hubble} Space Telescope (HST) and \textit{Spitzer} Space Telescope enabled us to probe the end of the EoR $(z\ga6)$ in great detail \citep[e.g.][]{ellis_abundance_2013, oesch_dearth_2018, atek_extreme_2018, bouwens_newly_2019} owing to very deep surveys such as the Hubble frontier field (HFF; \citealt{lotz_frontier_2017}). However, studies at higher redshift remained limited to the observations of a handful of very bright galaxies (z>8 objects; \citep[e.g.][]{zitrin_lyman_2015, oesch_remarkably_2016}) due to the wavelength coverage and limited sensitivity of HST, the resolution of \textit{Spitzer}, and the transmission of IR light for ground-based astronomy.
Since 2022, the \textit{James Webb} Space Telescope (JWST) enables us to probe farther and deeper into the EoR with regular detections of record redshift galaxies \citep[e.g.][]{curtis-lake_spectroscopic_2023, wang_uncover_2023, fujimoto_ceers_2023, harikane_pure_2024, napolitano_seven_2025, naidu_cosmic_2026} and studies of the faintest galaxies \citep[e.g.][]{eisenstein_overview_2023, bezanson_jwst_2024, suess_medium_2024, finkelstein_cosmic_2025}.

Uncovering the behaviour of the faintest galaxies (i.e. $\muv\geq-17$) is paramount to understanding the EoR. Faint galaxies are very numerous in the Universe, but remain difficult to study.
The star formation in these galaxies is highly unstable and characterised by periods of intense activity, followed by prolonged quiescence \citep[e.g.][]{endsley_burstiness_2025}.
On the one hand, the large abundance of extreme emission line emitters (EELGs), that is, of galaxies with emission lines reaching a rest-frame equivalent width of $\gtrsim 1000$Å, shows the signature of low-mass galaxies undergoing an intense star formation period \citep{rinaldi_midis_2023}. While rather rare at low redshift \citep{matthee_eiger_2023}, they appear to be more common at high redshift \citep[e.g.][]{atek_very_2011, wel_extreme_2011, smit_evidence_2014, tang_mmt_2019, izotov_diverse_2020, onodera_broadband_2020, berg_characterizing_2021, davis_census_2024, llerena_physical_2024, rinaldi_emergence_2025, boyett_extreme_2024, endsley_starforming_2024, begley_evolution_2025, daikuhara_nature_2025}. 
On the other hand, the numerous detections of galaxies with low star formation rates (SFRs), variously referred to as (mini-)quenched, in a phase of SFR downturn, or dormant \citep[e.g.][]{gelli_quiescent_2023, strait_extremely_2023, dome_mini_2024, endsley_burstiness_2025, looser_recently_2024, looser_jades_2025, mintz_taking_2025, trussler_like_2025, covelo-paz_systematic_2026}, shows the complex and bursty nature of star formation in the EoR, with short intense star formation periods followed by a longer period of very low star formation.
EELGs are more readily detectable as their strong emission lines boost the filters through which they are observed \citep{schaerer_impact_2009}.
Faint low-mass galaxies, with low star formation, lack this additional boost, which limits their detection. Their importance is further confirmed by recent studies, however, which showed evidence for missing low-mass galaxies in a statistical sample, with a low SFR, which skews the statistics \citep[e.g.][]{endsley_burstiness_2025, simmonds_ionizing_2024}.\\

The luminosity function (LF) characterises the number density of galaxies for a certain luminosity, providing a powerful description of the galaxy population demographics. Measuring the LF in different fields enables us to compare galaxy populations and examine variations in different regions. The UV LF was extensively studied in the EoR to understand the contribution of galaxies to the re-ionisation \citep[e.g.][Atek et al. in prep]{richard_hubble_2008, oesch_udf05_2009, oesch_dearth_2018, atek_extreme_2018, bouwens_alma_2020, moutard_uv_2020, bowler_lack_2020, bouwens_z_2022, willott_steep_2024, harikane_jwst_2025, weibel_exploring_2025, chemerynska_first_2026}.
Knowing the exact number density distribution enables us to assess the contribution of each type of galaxy by measuring their ionising photon-production rate \citep[e.g.][]{naidu_rapid_2020, atek_most_2024, simmonds_ionizing_2024}.
However, the challenges in constraining the Lyman-continuum escape fraction (\fesc) at high redshift means that this question remains unresolved.
\citet{jecmen_glimpse_2026} recently studied \fesc\ for the JWST GLIMPSE survey using its relation to $\beta$-slope \citep[e.g.][]{chisholm_far-ultraviolet_2022}. Their results indicated a constant $\fesc\sim14\%$, consistent with the $\fesc\sim10\%$ from earlier studies \citep{robertson_new_2013, robertson_cosmic_2015, giovinazzo_breaking_2025, mascia_little_2025}, but extends these constraints to significantly fainter galaxies. 
Emerging evidence, including from \citet{jecmen_glimpse_2026}, further suggests that \fesc\ might evolve with galaxy mass, thereby significantly affecting the relative contribution of faint and bright galaxies to re-ionisation \cite[e.g.][]{begley_vandels_2022, saldana-lopez_vandels_2023, jecmen_glimpse_2026}.
In addition, the ionising photon-production efficiency $\xi_{\rm ion}$ also requires attention. By overestimating it, the measured steep UV LF results in an overshooting of the ionising photon-production rate, and therefore, in fast re-ionisation \citep{robertson_new_2013, robertson_cosmic_2015, munoz_reionization_2024, simmonds_ionizing_2024, bosman_measurement_2024}.

The ionisation of the interstellar medium of galaxies produces a wealth of information through emission lines \citep[see][]{kewley_understanding_2019}. \ha~is a major tracer of the instantaneous star formation in galaxies \citep{kennicutt_star_2012}, but its observation with JWST/NIRCam is limited to redshift $z\leq6.5$. Therefore, we used the second-strongest Balmer series line \hb~to study star formation. However, due to the proximity of this line with \oiiifull~(hereafter \oiii), we were unable to distinguish them with broadband photometry, motivating the study of the combined \oiiihb\ complex. However, this adds complexity to tracing the star formation because the \oiii~doublet and \Rthree=\Oiiib-to-\hb\ ratio are heavily affected by the ISM metallicity \citep[e.g.][]{curti_new_2017, maiolino_re_2019, curti_mass-metallicity_2020, sanders_direct_2024, scholte_jwst_2025}.

The \oiii~or \oiiihb~LF has been measured multiple times for different fields \citep[e.g.][]{colbert_predicting_2013, khostovan_evolution_2015, de_barros_greats_2019, khostovan_large_2020, bowman_z_2021, matthee_eiger_2023, sun_first_2023, nagaraj_h_2023, meyer_jwst_2024, wold_uncovering_2025}, but was limited at high redshift by the wavelength coverage of WFC3/HST ($z\la3$), the resolution of \textit{Spitzer}/IRAC, and the poor infrared transmission through atmosphere for ground-based astronomy. The early results from \citep{de_barros_greats_2019} were able to push the redshift limits to $z \sim 8$ using a combination of \textit{Spitzer} $3.6\mu$m and $4.5\mu$m, the deep fields of HST, and an assumed relation between the UV LF and the \oiii~LF. 
\citet{meyer_jwst_2024} measured a non-biased \oiii~LF using the JWST/NIRCam/WFSS FRESCO \citep{oesch_jwst_2023} survey for relatively bright ($\muv\la-18$) galaxies between $6.8 < z < 9.0$. They found a number density lower than previously estimated and observed a rapid decline in the \oiii~number density at $z \gtrsim 7$. 
\citet{matthee_eiger_2023} measured the \oiii~LF at $5 < z < 7$ using JWST/NIRCam/WFSS EIGER \citep{kashino_eiger_2023} images and found little evolution of the \oiii~LF when compared to lower-redshift ranges ($z \lesssim 5$; \citealt{khostovan_evolution_2015}). 
However, until recently, the faint end of the LF was either indirectly constrained or simply fixed because only a few sources were detected. 
This is significant because the faint-end slope critically affects our understanding of the evolution of re-ionisation and the role of low-mass galaxies during this era \citep{naidu_rapid_2020, atek_most_2024, munoz_reionization_2024}. \citet{wold_uncovering_2025} was the first study to push the faint-end limits of the \oiii~LF back at high redshift. To achieve this, they used photometric data from the strongly lensed Abell 2744 HFF field, observed by the JWST ultradeep NIRSpec and NIRCam observations before the epoch of re-ionisation team \citep[UNCOVER, ][]{bezanson_jwst_2024}, which made use of magnification to obtain deeper observations. So far, \citet{wold_uncovering_2025} alone pushed the faint-end limits of the \oiii~LF back at high redshift ($L_{\Oiiib}\gtrsim10^{41}$\ergs\ for \citet{wold_uncovering_2025}, whereas $L_{\Oiiib}\gtrsim10^{41.75\ {\rm to}\ 42}$\ergs\ for \citet{matthee_eiger_2023, meyer_jwst_2024}.\\

To be able to reach the faintest galaxies and constrain the faint end of the \oiiihb~LF, we used the deepest JWST survey to date, GLIMPSE, on the strongly lensed HFF galaxy cluster Abell S1063. The unprecedented depth of the survey coupled to the strong magnification of some galaxies enabled us to reach the faintest galaxies ever observed at high redshift. We used a Lyman-break selection (Sect.~\ref{sec:selection}) and constrained the strong-lensing (SL) model to deduce magnification, multiplets, and effective volumes (Sect.~\ref{sec:sl} and \ref{sec:effective_vol}). We then made use of SED fits to measure their emission line fluxes (Sect.~\ref{sec:line_measurement}), we estimated the completeness of our sample (Sect.~\ref{sec:compl}), and we constructed the \oiiihb~LFs and analysed their behaviour (Sect.~\ref{sec:lf}). Finally, we discuss the implication of the resulting LFs with respect to the ionising photon budget required to re-ionise the Universe and the measured cosmic star formation rate density (SFRD; Sect~.\ref{sec:implication}). We considered a flat $\Lambda$CDM cosmology with $H_0 = 70$ \kmsmpc, $\Omega_M = 0.3$, and $\Omega_\Lambda = 0.7$. All magnitudes are expressed in the AB system \citep{oke_secondary_1983}, and we adopted the \citep{chabrier_galactic_2003} IMF.\\

\section{Data}

\subsection{Observations}
The GLIMPSE survey is a cycle 2 JWST/NIRCam large program (GO-3293; PI Atek \& Chisholm; \citealt{atek_jwst_2025}) that performed the deepest observations of the HFF (\citealt{lotz_frontier_2017}) galaxy cluster Abell S1063 ($z=0.348$). The observations consist of one pointing with two modules: the first one centred on the lensed field and the second one on a nearby region. The field was observed with 7 JWST/NIRCam broadband filters (F090W, F115W, F150W, F200W, F277W, F356W and F444W) reaching $5\sigma$ depths of $\sim 30.9$ mag and 2 JWST/NIRCam medium bands (F410M, F480M) reaching $5\sigma$ depth of $\sim30.1$ mag \citep{atek_jwst_2025}. In addition to GLIMPSE data, we also used shallow cycle 1 JWST/NIRCam observations with the F250M and F300M medium bands \citep{hashimoto_reionization_2023}, as these bands could provide some constraint on the SED fitting. In addition to the JWST/NIRCam observations, we also included the legacy data from the HST from the HFF survey \citep{lotz_frontier_2017} and the beyond ultra-deep frontier fields and legacy observations survey \citep[BUFFALO, ][]{steinhardt_buffalo_2020}. 
These observation provide deep imaging from HST/ACS (F606W, F814W) and HST/WFC3 (F105W, F125W, F140W, F160W), which constrain the rest-frame UV and blue-optical of the targeted galaxies in this study.\\

All the data were reduced following the procedure from \citep{endsley_starforming_2024}. In summary, the Point Spread Function (PSF) was built from the stars observed in the field. The JWST and HST images were then PSF matched to the reddest filter (F480M). The foreground bright cluster galaxies were subtracted following the method described in \citep{shipley_hff-deepspace_2018, weaver_uncover_2024, suess_medium_2024}. Then, we performed source extraction using \texttt{Sextractor} \citep{bertin_sextractor_1996} on an inverse-variance weighted stack of F277W+F356W+F444W, the broadband long wavelength filters of NIRCam. We extracted the photometry with an aperture size of D=0.2\arcsec and calculated the uncertainty by measuring the standard deviation from $2,000$ random photometric measurement at the same aperture in nearby empty regions. 
The small aperture size is motived by two key factors. First, the extreme depth of the GLIMPSE survey on the lensed field Abell S1063 results in a highly clustered field. Secondly, our primary science goal is the study of faint high-redshift galaxies, which are compact. Therefore, a smaller aperture ensures that we maximise the signal-to-noise for these galaxies, by reducing nearby contamination.
We corrected the aperture using empirical measurements of the PSF curve of growth, enabling us to measure the fraction of flux missing. The total flux is corrected from aperture by a factor 0.412. The full detail of the reduction is described in the GLIMPSE survey paper \citep{atek_jwst_2025}.

\subsection{Simulations}
In addition to observational data, we also used simulated fields to assess our results with simulations. We used the JWST extragalactic mock catalog \citep[JAGUAR,][]{williams_jwst_2018}, which is a catalogue of pre-JWST simulated galaxies with the aim of preparing for JWST science. These galaxies span a wide redshift range $0.2 < z < 15$ for galaxies of masses $\text{M}_* \geq 10^6\text{M}_\odot$. This simulation was mainly used to prepare for observations from the JWST advanced deep extragalactic survey \citep[JADES, ][]{rieke_jades_2023, eisenstein_overview_2023}. This simulation includes two branches: one containing star-forming galaxies and the other containing quiescent galaxies. In this paper, we only focus on the first realisation (R1) of the star-forming galaxies catalogue. To produce these catalogues, JAGUAR assumes an UV LF extrapolation from HST observations for its galaxy count, which can then be used to extrapolate the stellar mass of those galaxies. Galaxy models of BEAGLE \citep{chevallard_modelling_2016} were then used to generate realistic SEDs of the galaxies. These models span many parameters, which are then validated against observations of that time, such as 3D-HST \citep{skelton_3d-hst_2014}. 
As observations back then were limited in specific regions of the parameter space, such as low stellar mass galaxies ($\text{M}_* \leq 10^8\text{M}_\odot$), they relied on the theoretical knowledge of the time to forward-model galaxies.
In practice, they enforce a sharp cut at $M_* \geq 10^6 M_\odot$, which will affect our very faint galaxies comparison \citep{williams_jwst_2018}. However, galaxies with $\muv \leq -16$ should not be affected. While these simulations do not reproduce exactly current observations, they still enable us to test and calibrate our methods \citep[e.g.][]{meyer_jwst_2024, sun_first_2023}.\\

\section{Method and results}
\subsection{Source selection}
\label{sec:selection}

The selection of the present sample is based on a combination of the Lyman-break technique and photometric redshift estimates. The same approach is used by \citet{chemerynska_first_2026} at redshifts $z>9$. We first apply a colour-colour selection that aims at identifying the flux dropout in the bands blueward of rest-frame Lyman-$\alpha$, and at the same time minimise the contamination of low-redshift red sources by using the colours from the redder bands. For galaxies in the redshift range $7<z<9$, we relied on a dropout in the F090W filter by adopting the following colour-colour selection criteria:

\begin{equation}
        \begin{array}{l}
                m_{090}-m_{115}>0.8\\
                \land ~ m_{115}-m_{150}>1.6+(m_{150}-m_{200})\\
                \land ~ m_{115}-m_{150}<0.5 \lor m_{150}-m_{200}<0.5 \\
        \land ~ {\rm S/N}_{115}>5 \lor {\rm S/N}_{200}>5
        \end{array}
 \label{eq:selection},
\end{equation}

\noindent where we also required galaxies to be detected with high significance, at a $5\sigma$ level, in at least one band. In addition, we verify that all galaxies remain undetected at a $2\sigma$ level in all of the HST bands. Furthermore, we restricted our search area to regions with at least three-dither overlap to mitigate contamination by spurious sources. Accordingly, the survey area presented in Sect.~\ref{sec:effective_vol} has been corrected for these restrictions of the search area and we end up with an effective survey area of $4.30$ to $4.34$arcmin$^2$.

The second step of this selection is based on photometric redshifts computed with {\tt Eazy} \citep{brammer_eazy_2008}. We used the following parameters during the SED fitting procedure:  the {\tt blue\_sfhz\_13} templates, dust attenuation following \citep{calzetti_dust_2000} and IGM attenuation following \citep{inoue_updated_2014}. The {\tt Eazy} redshift grid spanned the full range, from $z=0.01$ to $z = 30$. We ensure that all dropout-selected galaxies have a best-fit solution at redshifts higher than $z>6$. The final sample contains 173 galaxies between $7<z<9$, and their \muv~distribution is shown in Fig.~\ref{fig:dist_Muv} ($\muv=-16.57_{-0.97}^{+1.63}$).

\subsection{Gravitational lensing} \label{sec:sl}

\begin{figure}
    \centering
    \includegraphics[width=\linewidth]{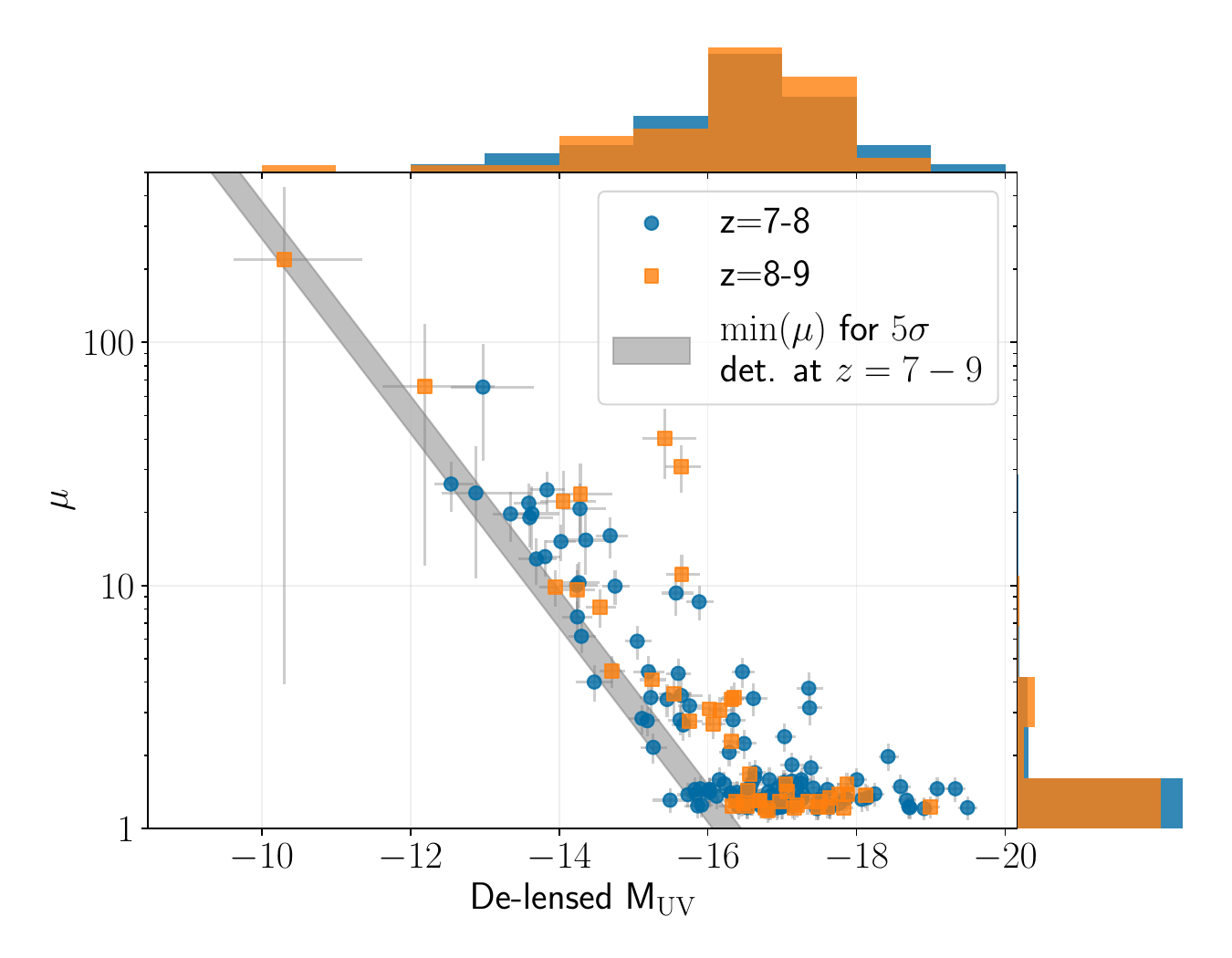}
    \caption{Relation between the de-lensed absolute UV magnitude \muv\ of the GLIMPSE F090W dropouts selected in this work and their SL magnification $\mu$. Beyond $M_{\mathrm{UV}}=-16$, we need magnification to detect these objects. This is shown by the grey area, which is the minimum magnification required for a $5\sigma$ detection at $z=7\text{ to }9$ with F115W, our deepest band \citep[see][]{atek_jwst_2025}. The top and right histograms show the normalised distribution of \muv\ and $\mu$ for each redshift bins.}
    \label{fig:dist_Muv}
\end{figure}

In order to account for the strong gravitational lensing (SL) effect of Abell~S1063, we use a new parametric SL mass model of the cluster, constructed with the \citet{zitrin_hubble_2015} parametric method \citep[see][]{furtak_uncovering_2023}, revised to be fully analytic, i.e.\ not limited to a grid resolution. The lens model will be presented in detail in Furtak et al. (in prep.) and we refer the reader to that work for more details.

In short, the model comprises two smooth dark matter (DM) halos parametrised as pseudo-isothermal elliptical mass distribution (PIEMD; \citealt{kassiola_elliptic_1993}), one centred on the brightest cluster galaxy (BCG), and the other on a group of galaxies in the north-east of the cluster as found by previous works \citep[][]{bergamini_enhanced_2019,beauchesne_new_2024}. In addition, the model includes 303 cluster member galaxies, parametrised as dual pseudo-isothermal ellipsoids (dPIEs; \citealt{eliasdottir_where_2007}). The model is constrained with 75 multiple images of 28 sources, 24 of which have spectroscopic redshifts. Optimised using a Monte Carlo Markov chain (MCMC) analysis, the lens model achieves a final lens-plane image reproduction RMS of $\Delta_{\mathrm{RMS}}=0.54\arcsec$. This model has previously been used various projects \citep[e.g.][]{topping_metal-poor_2024, kokorev_glimpse_2025, fujimoto_glimpse_2025, chemerynska_first_2026}.

We use the lens model to compute the gravitational magnification at the coordinates and redshift of each objects. All sources, even in the off-cluster module, are magnified by at least a factor $\mu=1.19$. In Fig.~\ref{fig:dist_Muv}, we show the explored \muv~against magnification $\mu$. We see that for $\muv \geq -16$, stronger magnifications are required to even detect the objects, reaching $\max(\mu) \sim 219$ for our strongest-lensed galaxy. 

We also need to account for multiple imaging in our object counts when deriving the LF in order to avoid double- or even triple-counting galaxies. This is done in an iterative process where image positions are injected into the lens model to predict eventual counter image positions. We then search the data for corresponding sources in the 3 arcseconds area surrounding the predicted counter-images location and carefully matched photometries and SEDs to make sure they are the same object. In the event of multiple imaging, we keep the counter-image with the highest F444W signal-to-noise. From our sample of 173 galaxies, 68 expect one or more counter-images. 9 of them can be found within our Lyman-Break Galaxy LBG sample at 7<z<9, which reduces our total sample size to 164 galaxies.

\subsection{Estimating the effective survey volume}
\label{sec:effective_vol}
Estimating the effective volume of each source relies on our SL model of the field. As described in \citep{atek_extreme_2018}, the effective volume is estimated as the integral of the co-moving volume that comes from the cosmology to which we apply completeness and magnification. It corresponds to the source plane volume in which we effectively observe a galaxy of observed magnitude $m$. Typically, a faint source can only be detected in highly magnified regions, which reduces the effective volume for these sources. The effective volume can be analytically computed and is defined as

\begin{equation}
    \text{V}_\text{eff} = \int_0^\infty dz\int_{\mu>\mu_{min}}d\Omega(\mu, z) \frac{\text{V}_\text{com}}{dz} C(z, m, \mu) \approx V \times C
    \label{eq:veff},
\end{equation}

\noindent where $\text{V}_\text{eff}$ is the effective volume, $\text{V}_\text{com}$ is the co-moving cosmological volume, $C(z, m, \mu)$ the detection completeness for sources at redshift $z$, with apparent magnitude $m$ and magnification $\mu$, and $d\Omega(\mu, z)$ is the surface element that depends on magnification and redshift \citep{atek_extreme_2018}. The volume uncertainties are obtained by propagating the magnification-dependent systematic error in the cumulated source plane area, the error on the completeness described in Sect. \ref{sec:compl}, and the variation of the effective volume in the \muv~bin. In this work, we extracted the completeness out of the integral by considering discrete bins of completeness and volumes instead of continuous functions. We refer to $V$ as the average volume and to $C$ as the average completeness over a given \muv~bin.

\begin{figure}[t]
    \centering
    \includegraphics[width=\linewidth]{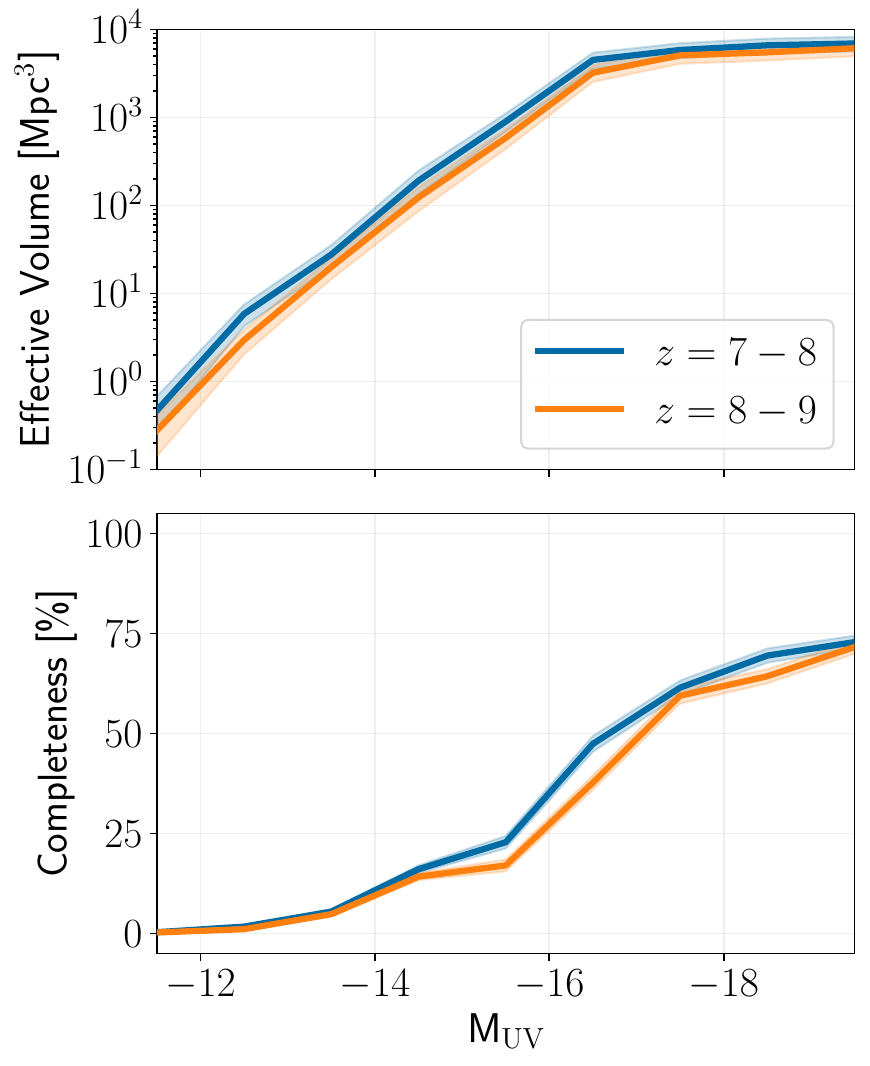}
    \caption{Top panel: Evolution of the effective volume probed for each UV magnitude M$_\text{UV}$ for our two redshift bins 7<z<8 (blue) and 8<z<9 (orange) with 1$\sigma$ uncertainties in the shaded area. Bottom panel: Completeness estimate for the same \muv\ and redshift bins.}
    \label{fig:veff}
\end{figure}

Figure~\ref{fig:veff} shows the evolution of the effective volume with respect to the galaxy UV magnitude M$_\text{UV}$. For brighter sources ($\text{M}_\text{UV} < -17$), a plateau is reached that corresponds to sources that should be detected in the entire image regardless of magnification, but for fainter sources ($\text{M}_\text{UV} > -17$), the magnification becomes important for the detection, which then quickly reduces the surveyed effective volume. \\

In summary, we integrated the source planes from the SL model for sources at different redshifts, with different apparent magnitudes and located in different magnification regions to obtain the effective volumes for each galaxy.

\subsection{Line flux measurement of \oiiihb}
\label{sec:line_measurement}
We measured line fluxes by fitting the SEDs of our dropout-selected galaxies. For this, we use the CIGALE \citep{boquien_cigale_2019} SED fitting software with the fixed redshift obtained by EASY. We used a delayed-$\tau$ (\texttt{sfhdelayed}) star formation history with an e-folding time and an age interval between 1 Myr and 13.5 Gyr. Galaxies are bound to the redshift, and therefore, no galaxies exceed 500 Myr. We used the Bruzual\&Charlot (\texttt{bc03}) \citep{bruzual_stellar_2003} simple stellar populations assuming a Chabrier initial mass function (IMF; \citealt{chabrier_galactic_2003}) and a fixed gas and stellar metallicity of 0.004 (the mass fraction of atoms heavier than helium, where $Z_\odot = 0.02$). We include nebular templates with the ionisation parameter varying between $\log U =-1$ and $-4$ and an electron density of 100 cm$^{-3}$. We also account for dust attenuation, using a module based on a modification by \citep{leitherer_global_2002} of the law from \citep{calzetti_dust_2000}. We vary the $E(B-V)_{l}$ colour excess of the nebular lines between 0 and 1.5 and a power law modifying the attenuation curve between -0.5 and 0. In addition, we also adopt the Milky Way extinction curve from \citep{cardelli_relationship_1989} updated by \citep{odonnell_r_1994} to better match high resolution extinction curves from \citep{bastiaansen_narrow_1992}. More details on the different fitting modules are described in \citep{boquien_cigale_2019}. We measure the line flux using the internal logic from CIGALE, which uses a Bayesian estimation for each measured parameter by weighting templates by their likelihood. In CIGALE, the gas and stellar metallicity are considered as two separate quantities. Because of their degeneracies with other parameters, they end up being poorly constrained and might yield unphysical solutions with large differences between stellar and gas metallicities. Therefore, we decided to keep the metallicity fixed to $Z=0.004$ in the fitting. As a test, we also varied the metallicity, but no significant differences in the LFs were observed. Therefore, for the performance improvement as well as coherence of the gas and stellar metallicity, we decided to keep the SED fitting metallicities parameters constant. Three representative galaxies with stamps cutouts and SED can be found in Appendix\ \ref{app:sed_examples}.\\

To validate our measurement method, we assessed it with the JAGUAR simulations \citep{williams_jwst_2018}, which showed a good correlation (correlation coefficients of 0.93 and 0.79) between the absolute JAGUAR value of \oiiihb~flux and EW and the retrieved SED fitting estimation of the same parameters. In addition, we also validated our method through a direct measurement of the \oiiihb~flux and EW based on excess in the available F356W, F410M, F444W and F480M bands, which again showed a reasonably good correlation coefficient of 0.67 for the flux, but 0.20 for the EW. We see some divergence on the faint end for the EW, but this most probably comes from greater uncertainties in the direct measurement due to a lack of medium band availability. The details of the validation can be found in Appendix~\ref{app:validation}.

\subsection{Dust attenuation}
\label{sec:dust_att}
\begin{figure}[t]
    \centering
    \includegraphics[width=\linewidth]{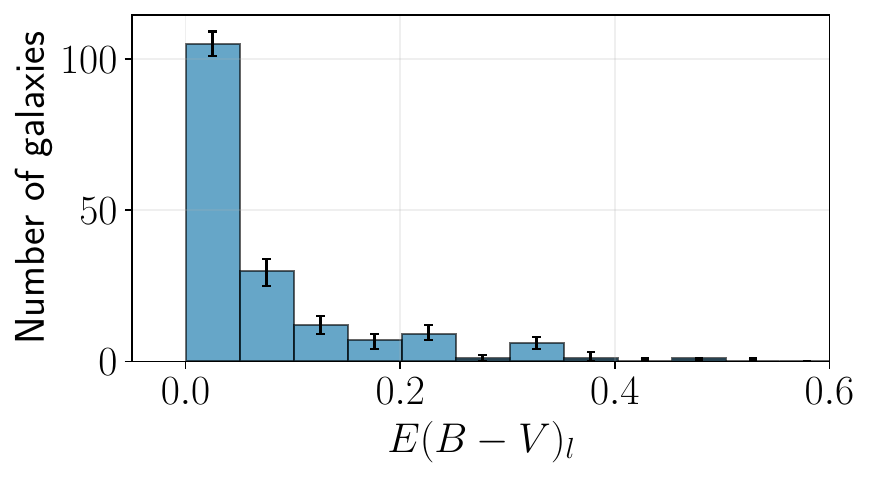}
    \caption{Histogram of the dust attenuation parameter $E(B-V)_l$ sampled from the SED fitting of our sample of $7<z<9$ galaxies. $N=10\,000$ realisations of the measurement are performed for each galaxy, following the distributions of the SED fitting, and the histogram is normalised by $N$.}
    \label{fig:dust_attenuation_hist}
\end{figure}

Unless otherwise stated, the results reported in this paper correspond to observed quantities, which are not corrected for dust attenuation. To correct for attenuation, we use SED fitting, as described in Sect.~\ref{sec:line_measurement}, to infer the line colour excess $E(B-V)_l$ which is defined in \citet{calzetti_dust_2000} as the continuum $E(B-V)$ divided by an empirical factor $0.44\pm0.03$. The inferred emission line fluxes (or luminosities) are then corrected for dust attenuation, adopting the extinction law from \citep{cardelli_relationship_1989}. Figure \ref{fig:dust_attenuation_hist} shows the $E(B-V)_l$ distribution of our sample. The median measurement lies at $E(B-V) = 0.013$, and the majority of the galaxies are compatible with a low attenuation of $E(B-V)_l < 0.1$. We do not find significant correlations between dust attenuation and parameters such as \muv. Therefore, these low $E(B-V)_l$ values imply that dust is unlikely to play an outsized role in the inferred \oiiihb~flux values.

\subsection{\oiiihb~completeness estimation}
\label{sec:compl}
\begin{figure}[t]
    \centering
    \includegraphics[width=1\linewidth]{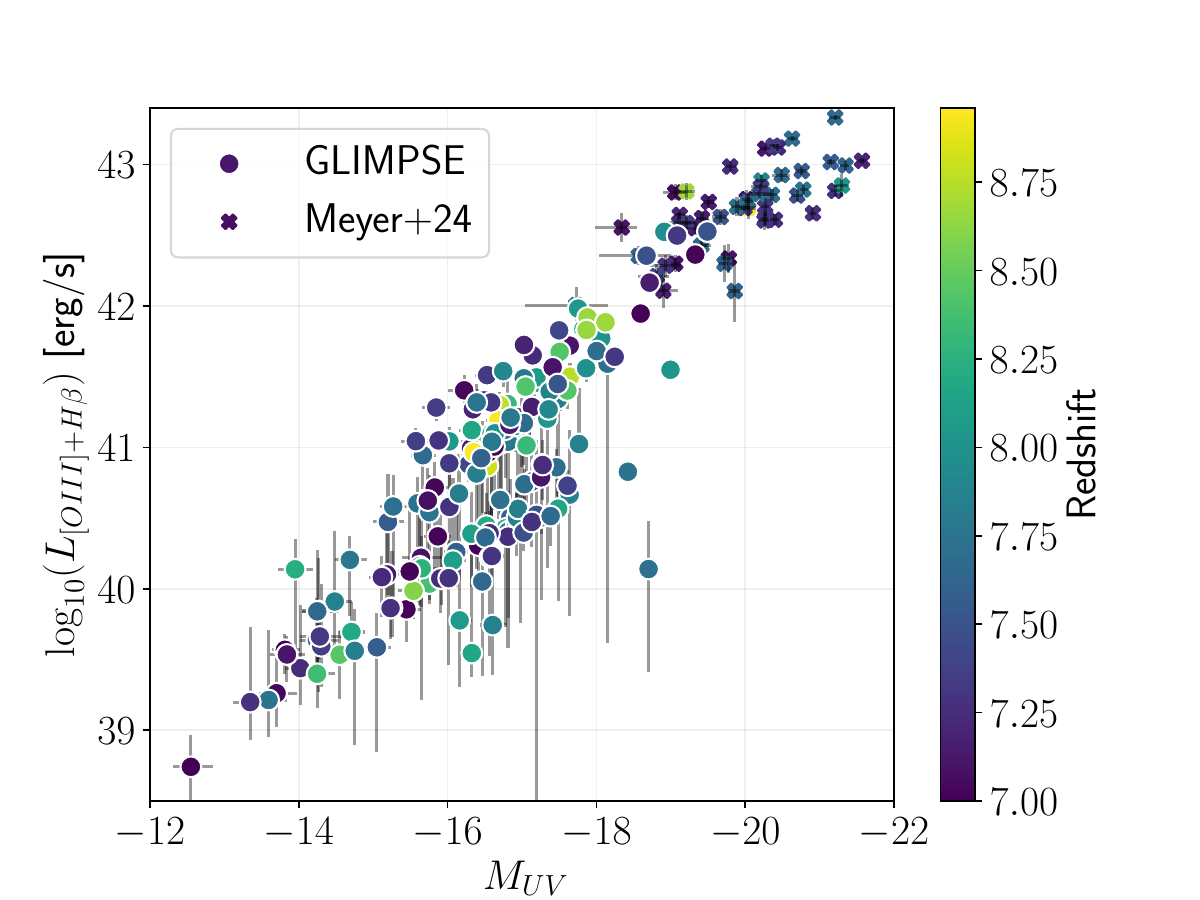}
    \caption{Correlation between the \oiiihb~line flux against $M_{UV}$ for our GLIMPSE sample (circles) and measurements from \citet[][shown as crosses]{meyer_jwst_2024}. To match our observations, we only kept sources within the same $7 < z < 9$ redshift range. Bayesian error estimates using CIGALE are also shown for the GLIMPSE measurements.}
    \label{fig:muv_flux}
\end{figure}

Given the non-trivial nature of oxygen lines, we used an indirect method to estimate the completeness of our sample via $M_{UV}$. We used the completeness estimation made by the GLIMPSE collaboration \citep[e.g.][]{chemerynska_first_2026}, which follows the procedures described in \citep{chemerynska_jwst_2024, atek_extreme_2018}. In summary, synthetic galaxies were added into the source plane, mapped into the lens plane using the SL model (see Sect.~\ref{sec:sl}), added to the images, and then selected using the same process as described earlier. This \muv-based completeness does not exactly correspond to our \oiii~completeness. However, \muv\ and the \oiiihb~line flux correlate \citep{matthee_eiger_2023}. Figure~\ref{fig:muv_flux} shows the correlation between the \oiiihb~line flux and the \muv~for the GLIMPSE sample and the sample from \citet{meyer_jwst_2024}. The sample shows a good correlation between the two quantities, but shows a bimodal distribution of $L_{\oiiihb}$ for $\muv > -17$. This bimodality emerges from the poorer constrains on the faintest observed objects. At this magnitude, the galaxies are near the observed detection limit of GLIMPSE and their signal-to-noise is limited. Therefore, degeneracies between fitting parameters increase and the flux estimation becomes uncertain as many model fit the data. These galaxies are then propagated to lower magnitude \muv\ due to magnification, which gives this second mode in the \oiiihb-to-UV relation. However, while their flux measurements are more uncertain, the values inferred from the model are physically motivated and the posterior distributions will be sampled to account for their variety. Therefore, despite the scatter, the quantities correlate well, and the completeness for \muv~approximately translates to \oiiihb~line flux.\\

To measure uncertainties for the completeness, we assumed that it follows a binomial distribution $B(n, p)$ where $n$ is the total number of synthetic sources and $p$ is the completeness for this type of source (depending on the $M_{UV}$ and magnification $\mu$ of the source). This yields the estimation of the standard deviation for the completeness $\sigma_C$, with $C$ the completeness, as given by 

\begin{equation}
        \delta C = \sqrt{\frac{C(1-C)}{n}}
        \label{eq:std_completeness}.
\end{equation}

\section{The \oiiihb~luminosity function at $z \sim 7-9$}
\label{sec:lf}
We now describe our standard approach to measuring the \oiiihb~LF. 
We computed the LF using the following equation:

\begin{equation}
    \Phi(L)d\log_{10}(L) = \sum_i \frac{1}{V_i C_i}
    \label{eq:def_lf},
\end{equation}

\noindent where $L$ is the \oiiihb~line luminosity, $V_{i}$ is the effective survey volume for a given source $i$, $C_i$ is its associated completeness at given redshift and $L$.

To construct the LF, we first varied the line flux \oiiihb~(and \muv~for the completeness correction) from the measurements of the fitted SED templates. For this, we extracted the measurements of both quantities from CIGALE, as well as the associated template $\chi^2$. By weighting each measurement by the likelihood $\mathcal{L}=\exp(-0.5\chi^2)$, we obtain a 2D histogram (N-dimensions when dust attenuation or other parameters are sampled) from which we sampled $n = 10,000$ realisations of the parameters. To avoid discretisation effects caused by the sampling of the histogram, we uniformly varied the measurements within their bins. The \oiiihb~line flux was treated on a logarithmic scale. As the SED was fitted to the observed photometry, the magnification was also sampled following a normal distribution, and was applied to \muv~and \oiiihb~flux.\\

From there, we measure the LF and its associated uncertainties from each realisation. While the LF is measured following Eq. \eqref{eq:def_lf}, the uncertainties are more complex and come from multiple sources. First, we measured the intrinsic error of each luminosity bin coming from the uncertainties of the effective survey volume (which includes the completeness). By propagating uncertainties, we end up with the following:  

\begin{equation}
    \delta_\text{I}\log_{10}\Phi(L) = \frac{\sqrt{\sum_i \left(\frac{\delta V_i}{V_i^2C_i}\right)^2 + \left(\frac{\delta C_i}{V_iC_i^2}\right)^2 }}{\ln(10) d\log_{10}(L) \sum_i \frac{1}{V_i C_i}}
        \label{eq:def_lf_err},
\end{equation}

\noindent where $\delta_\text{I}\log_{10}\Phi(L)$ is the intrinsic uncertainty on the log-LF, $\delta V_i$ is the uncertainty on the volume and $\delta C_i$ is the uncertainty on the completeness.\\

In addition to the intrinsic errors, we also account for Poissonian errors given by the number of sources in each bin of the LF. For a bin uncertainty given by $\sqrt{N}$, with $N$ the number of sources in each luminosity bin, we obtain the following:

\begin{equation}
        \delta_\text{P}\log_{10}\Phi(L) = \frac{1}{\sqrt{N}\ln(10)d\log_{10}(L)}
        \label{eq:lf_poisson_err}
.\end{equation}

The intrinsic error and the Poissonian error are then summed in quadrature to obtain the final measurement of the LF for each realisation. We therefore have $n=10,000$ realisations of the same LF, where each luminosity bin has a measurement with uncertainties. To combine them all into one final LF, we sample $100$ subsample of each LF realisation, giving us $10,000 \times 100$ LF for which we measure mean and standard deviation.

In order to reduce biases from very low statistics due to non-detection in the faint end and small volume in the bright end, we applied two cuts to the LFs. First, each luminosity bin needs to have at least one source on average ($\mean{N}\geq1$), meaning that out of all realisations, on average, at least one object should lie in the luminosity bin. To avoid cases with large uncertainties on the \oiiihb~line flux, we measure the median value for each galaxy and we further enforced that each \oiiihb~bin needs at least one median object ($N\geq1$). 

While we use a Bayesian approach for our statistics, we would like to estimate the quality of our results using some signal-to-noise ratio (S/N) estimation. For this, we assume a log-normal distribution of our luminosity bins and use the general definition of the expected value ($E$) and the variance (Var) to measure our signal-to-noise. Then, using $\mu$ and $\sigma$ as the bins mean and standard deviation, we estimate the following equations:
\begin{align}
    E &= \exp(\mu + \frac{\sigma^2}{2})\\
    {\rm Var} &= (e^{\sigma^2} - 1) \exp(2\mu + \sigma^2)\\
    {\rm S/N} &= \frac{E}{\sqrt{\rm Var}}.
\end{align}
The S/N of each bin is reported in their associated tables.

\begin{figure*}[t]
    \centering
    \includegraphics[width=0.8\linewidth]{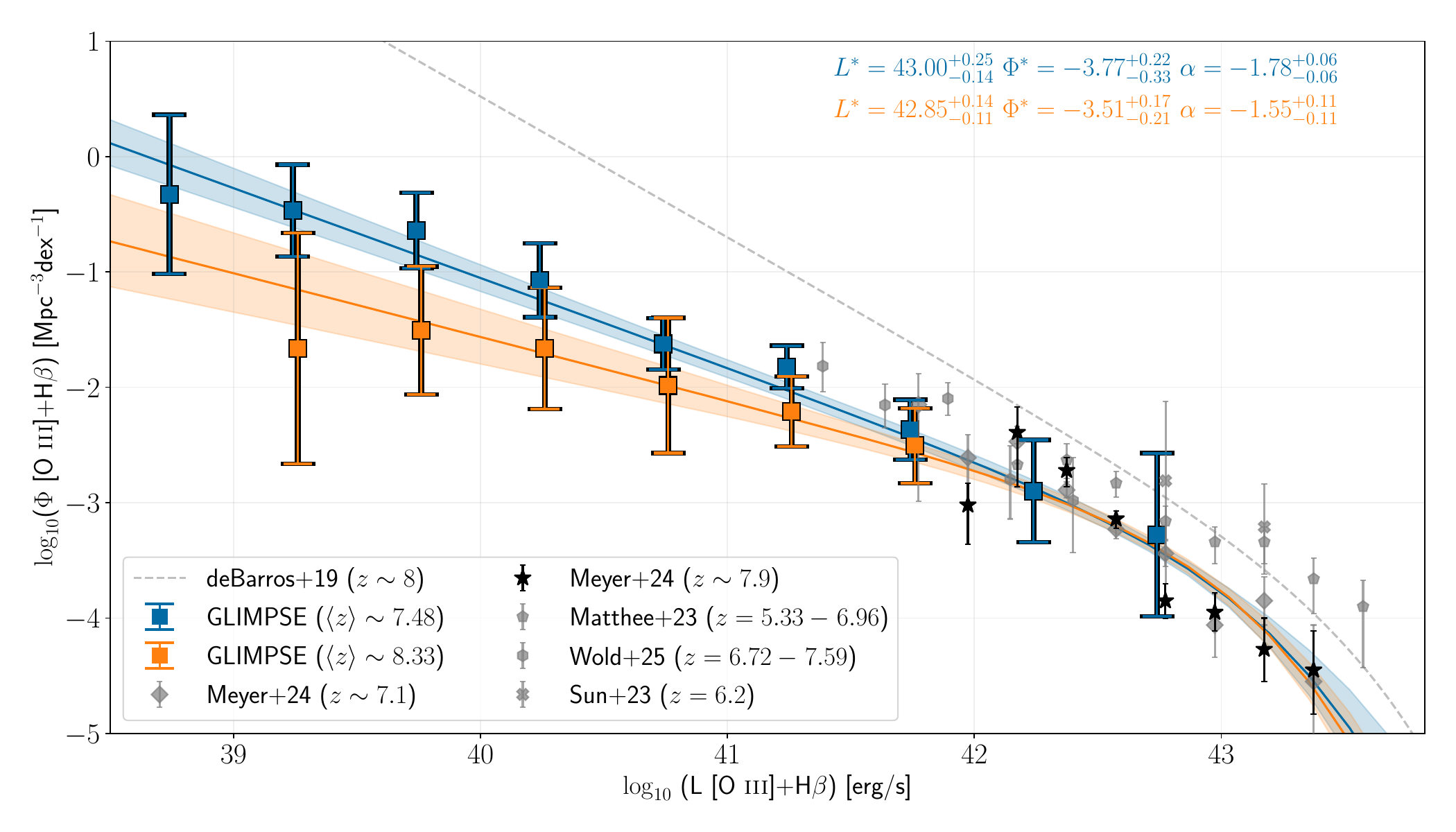}
    \caption{Luminosity function of \oiiihbfull~for our two redshift ranges. The GLIMPSE data points give the mean and standard deviation of the LF, while the fits provide the median value as a solid line, and the 16-84 percentile in the shaded area. We added previous studies of the LF using JWST/NIRCam GRISM slitless spectroscopy \citep{meyer_jwst_2024, matthee_eiger_2023, sun_first_2023}, a JWST/NIRCam medium band survey by \citet{wold_uncovering_2025} and a former \textit{Spitzer} study by \citet{de_barros_greats_2019}. All the JWST surveys specifically study the \oiii$\lambda5007$Å, so we converted them to \oiiihb~by considering their respective measured R3 factor as well as $\Oiiib/\Oiiit = 2.98$ \citep{storey_theoretical_2000}. The data can be found in Table~\ref{tab:O3LF} and the parametrisation can be found in Table~\ref{tab:schechter}.}
    \label{fig:O3LF}
\end{figure*}

\subsection{The \oiiihb~luminosity function at $z\sim7-9$}
\label{sec:behaviour}
In Fig.~\ref{fig:O3LF} we show the \oiiihb~LF for our two redshift ranges, and their comparison to the values from literature. 
We separated our LF in luminosity bins of 0.5 dex, resulting in 9 bins for the $z=7-8$ ($\mean{z}\sim 7.48$) redshift range and 6 bins for the $z=8-9$ ($\mean{z}\sim 8.33$) redshift range. We observe large uncertainties on the bright and faint-ends of the LF. On the bright end, the low number statistics drives the Poissonian errors, but on the faint-end, the addition of the uncertainties caused by degeneracies between the different solutions in the SED fitting models and the small effective volume increases the uncertainties. Nevertheless, we have a relatively low uncertainty between $10^{40}$\ergs and $10^{42}$ \ergs, with ${\rm S/N}>3$. The strength of GLIMPSE is the depth reached. Indeed, by combining a photometry approach on a very deep lensed galaxy field, we are able to reach unprecedented \oiiihb~line fluxes. The deepest comparison from \citet{wold_uncovering_2025} allowed us to securely probe the \oiiihb~LF down to $\sim10^{41.5}$ \ergs, while GLIMPSE goes down to $\sim 10^{39}$ \ergs, which constrains the high redshift faint end of the \oiiihb~LF to unprecedented depths. The full detail of the \oiiihb~LF is provided in Table~\ref{tab:O3LF}.

Because most of the previous studies only computed the \Oiiib~emission line, we needed to convert these previous measurements into \Oiiihb. To do this, we assumed an oxygen line ratio of $\Oiiib/\Oiiit = 2.98$ \citep{storey_theoretical_2000} and the \Rthree=\Oiiib/\hb~ratio measured or assumed in each study ($\Rt=6.72$ in \citet{wold_uncovering_2025}, 6.38 in \citet{meyer_jwst_2024}, 6.30 in \citet{matthee_eiger_2023}, 6.3 in \citet{sun_first_2023}, and \citet{de_barros_greats_2019} report directly \oiiihb). 
While this approach simplifies the conversion, since a rigorous treatment requires refitting each LF with individual \Rt\ measurements, we adopt it to constrain the bright end of our LF, using the results from \citet{meyer_jwst_2024} as reference.
The agreement between our LF and previous literature results in Fig.~\ref{fig:O3LF} suggests that this simplification does not significantly bias our results.

To parametrise our LFs, we use the common Schechter function \citep{schechter_analytic_1976} as given by

\begin{equation}
    \phi(L)dL = \phi^*\left(\frac{L}{L^*}\right)^\alpha\exp\left(-\frac{L}{L^*}\right)d\left(\frac{L}{L^*}\right)
    \label{eq:schechter}.
\end{equation}

This function is parametrised by three variables: the normalisation $\phi^*$, the characteristic luminosity $L^*$ and the faint-end slope $\alpha$. While we can probe very deep into the \oiiihb~LF, the field area (one NIRCam pointing in two modules) strongly limits the ability to probe brighter objects, which are less common than their faint counterparts. Indeed, we do not detect objects above $10^{43}$ \ergs\ and only a handful of objects above $10^{42}$ \ergs. Given that the characteristic luminosity is typically located in the vicinity of these luminosities, GLIMPSE does not enable us to strongly constrain the bright end of the \oiiihb~LF. Therefore, we combine our results with those of \cite{meyer_jwst_2024}, who studied the bright end of the \Oiiib~LF using an unbiased sample of GRISM spectra. They measured the LF down to $10^{41.75}$ \ergs\ for the high-redshift ($\mean{z}\sim7.9$) sample, which is at a redshift comparable to our lower-redshift sample. Their \Oiiib~LF is converted to \oiiihb~thanks to the median $\Rthree=\Oiiib/\hb=6.38 \pm 0.85$ measurement from their stacked spectroscopic spectra, enabling us to securely constrain the bright end of the LF, while leaving the faint-end constraints to GLIMPSE.

We used a MCMC approach to fit the LF. 
The likelihood $\mathcal{L}$ that we use is defined as 

\begin{equation}
\begin{split}
    \log_{10}\mathcal{L} = -\frac{1}{2} \Bigg[ & \sum_{L}^\text{GLIMPSE} \left(\frac{\phi(L) - \Phi(L,\phi^*, L^*, \alpha)}{\sigma_\phi(L)}\right)^2 \\
 + & \sum_{L}^\text{FRESCO} \left(\frac{\phi(L) - \Phi(L,\phi^*, L^*, \alpha)}{\sigma_\phi(L)}\right)^2 \Bigg]
    \label{eq:likelihood}
\end{split},
\end{equation}

\noindent where $\phi(L)$ is the mean observed density with uncertainty $\sigma_\phi(L)$ for a given luminosity $L$, $\Phi(L,\phi^*, L^*, \alpha)$ is the estimation of the parametrisation of the Schechter function at a given luminosity $L$. This parametrisation $\Phi$ corresponds to the convolution of the Schechter function with a Gaussian kernel, allowing us to account for the Eddington bias \citep{eddington_formula_1913}, which is a selection bias in which rare categories of objects are more often contaminated with more common categories of objects than the reverse because of their abundance. Finally, the first summation spans the luminosity range from GLIMPSE, and the second one from FRESCO \citep{meyer_jwst_2024} at $\mean{z}\sim7.9$. 
To estimate the uncertainties on the Schechter parameters, we sampled $n=10,000$ realisations of the LF and fitted them with the above procedure. We then measured the median and 16-84\% percentiles. 
The resulting fit parameters are listed in Table \ref{tab:schechter}.

\begin{table}[t]
    \caption{Schechter parametrisation of the LFs.}
    \centering
    \begin{tabular}{c|ccc}
        \hline\hline
        Redshift bin & $L^*$ & $\phi^*$ & $\alpha$ \\
        \hline

        \multicolumn{4}{c}{Fig.~\ref{fig:O3LF}: \oiiihb~LF}\\
        $\mean{z}\sim7.48$ & $43.00_{-0.14}^{+0.25}$ & $-3.77_{-0.33}^{+0.22}$ & $-1.78_{-0.06}^{+0.06}$ \\
        $\mean{z}\sim8.33$ & $42.85_{-0.11}^{+0.14}$ & $-3.51_{-0.21}^{+0.17}$ & $-1.55_{-0.11}^{+0.11}$ \\

        \multicolumn{4}{c}{}\\
        \multicolumn{4}{c}{Fig.~\ref{fig:O3LF_woldlike}: \oiiihb~LF limited at $10^{41}$\ergs}\\
        $\mean{z}\sim7.48$ & $43.88_{-0.78}^{+4.17}$ & $-5.23_{-5.39}^{+1.20}$ & $-2.19_{-0.12}^{+0.16}$ \\
        $\mean{z}\sim8.33$ & $43.15_{-0.27}^{+1.10}$ & $-4.02_{-1.53}^{+0.46}$ & $-1.92_{-0.25}^{+0.24}$ \\

        \multicolumn{4}{c}{}\\
        \multicolumn{4}{c}{Fig.~\ref{fig:HbLF}: \hb~LF}\\
        $\mean{z}\sim7.48$ & $42.17_{-0.21}^{+0.58}$ & $-4.03_{-0.76}^{+0.33}$ & $-1.95_{-0.08}^{+0.08}$ \\
        $\mean{z}\sim8.33$ & $41.96_{-0.14}^{+0.28}$ & $-3.65_{-0.40}^{+0.22}$ & $-1.68_{-0.14}^{+0.13}$ \\
        
        \multicolumn{4}{c}{}\\
        \multicolumn{4}{c}{Fig.~\ref{fig:O35007LF}: \Oiiib~LF}\\
        $\mean{z}\sim7.48$ & $42.78_{-0.13}^{+0.17}$ & $-3.68_{-0.22}^{+0.18}$ & $-1.66_{-0.05}^{+0.05}$ \\
        $\mean{z}\sim8.33$ & $42.65_{-0.10}^{+0.13}$ & $-3.45_{-0.17}^{+0.16}$ & $-1.45_{-0.10}^{+0.09}$ \\

        \multicolumn{4}{c}{}\\
        \multicolumn{4}{c}{Fig.~\ref{fig:HaLF}: dust-corrected \ha~LF}\\
        $\mean{z}\sim7.48$ & $42.60_{-0.19}^{+0.46}$ & $-3.99_{-0.63}^{+0.29}$ & $-1.96_{-0.08}^{+0.08}$ \\
        $\mean{z}\sim8.33$ & $42.38_{-0.13}^{+0.17}$ & $-3.60_{-0.26}^{+0.21}$ & $-1.67_{-0.14}^{+0.14}$ \\
        \hline\hline
        
    \end{tabular}
    \label{tab:schechter}
\end{table}

\subsection{The characteristic luminosity and normalisation parameters}
At the bright end of the LF, we obtain a characteristic luminosity of $L^* = 43.00_{-0.14}^{+0.25}$ and $L^* = 42.85_{-0.11}^{+0.14}$ for the $\mean{z}\sim7.48$ and $\mean{z}\sim8.33$ LFs respectfully. It agrees well with the literature results at similar redshifts \citep[e.g.][]{meyer_jwst_2024, wold_uncovering_2025, de_barros_greats_2019} but differs from lower redshift studies, which indicates some evolution with redshift \citep[e.g.][]{khostovan_evolution_2015, matthee_eiger_2023}. However, it is difficult to assess the impact of GLIMPSE in constraining the characteristic luminosity $L^*$ because of the very limited sample. The uncertainties of GLIMPSE are large in the bright end, and the result is mostly driven by \citet{meyer_jwst_2024}.

\begin{figure}[h]
    \centering
    \includegraphics[width=\linewidth]{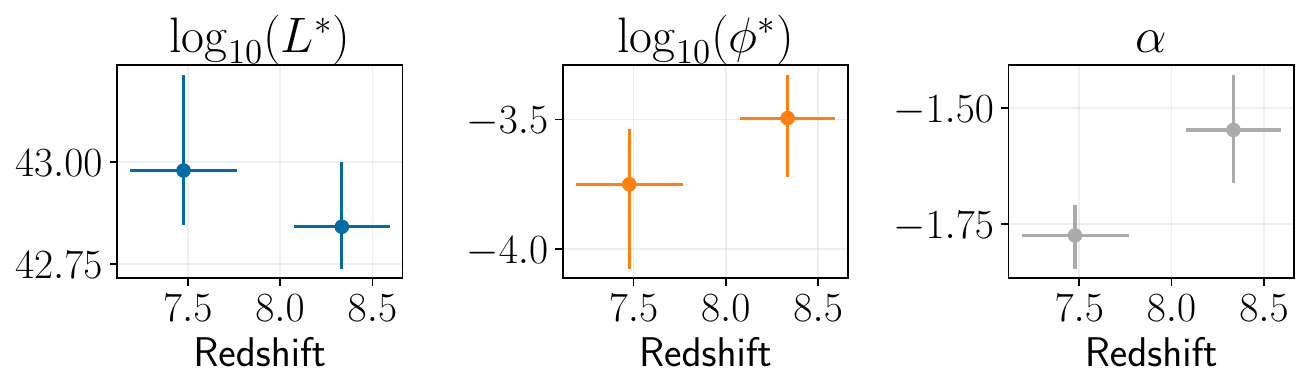}
    \caption{Redshift evolution of the fitting parameters from Fig.~\ref{fig:O3LF}.}
    \label{fig:par_zevo}
\end{figure}

Regarding the normalisation, we obtain $\phi^* = -3.77_{-0.33}^{+0.22}$ and $\phi^* = -3.51_{-0.21}^{+0.17}$ for the $\mean{z}\sim7.48$ and $\mean{z}\sim8.33$ LFs respectfully. It agrees within uncertainties with studies at similar redshift \citep[e.g.][]{meyer_jwst_2024, wold_uncovering_2025}, except for \citet{de_barros_greats_2019} which quickly diverge from our observations. However, this last study did not measure \oiiihb~directly, but converted it from UV using a UV-to-\oiiihb~calibration with \textit{Spitzer} and HST data, which can explain the visible difference. For lower redshifts, there seems to be an evolution between $z=0-3$ with a decreasing normalisation with increasing redshift \citep{colbert_predicting_2013, khostovan_evolution_2015, khostovan_large_2020, bowman_z_2021, nagaraj_h_2023}, but little evolution between $z=3-9$ \citep[e.g.][]{khostovan_evolution_2015, matthee_eiger_2023}. GLIMPSE further confirms that trend with no statistically significant evolution between redshift 7 to 9. The evolution of all three fitting parameters is shown in Fig.~\ref{fig:par_zevo}.

\subsection{Faint end of the \oiiihb~luminosity function}
\label{sec:faint_end}

\begin{figure}[tb]
    \centering
    \includegraphics[width=\linewidth]{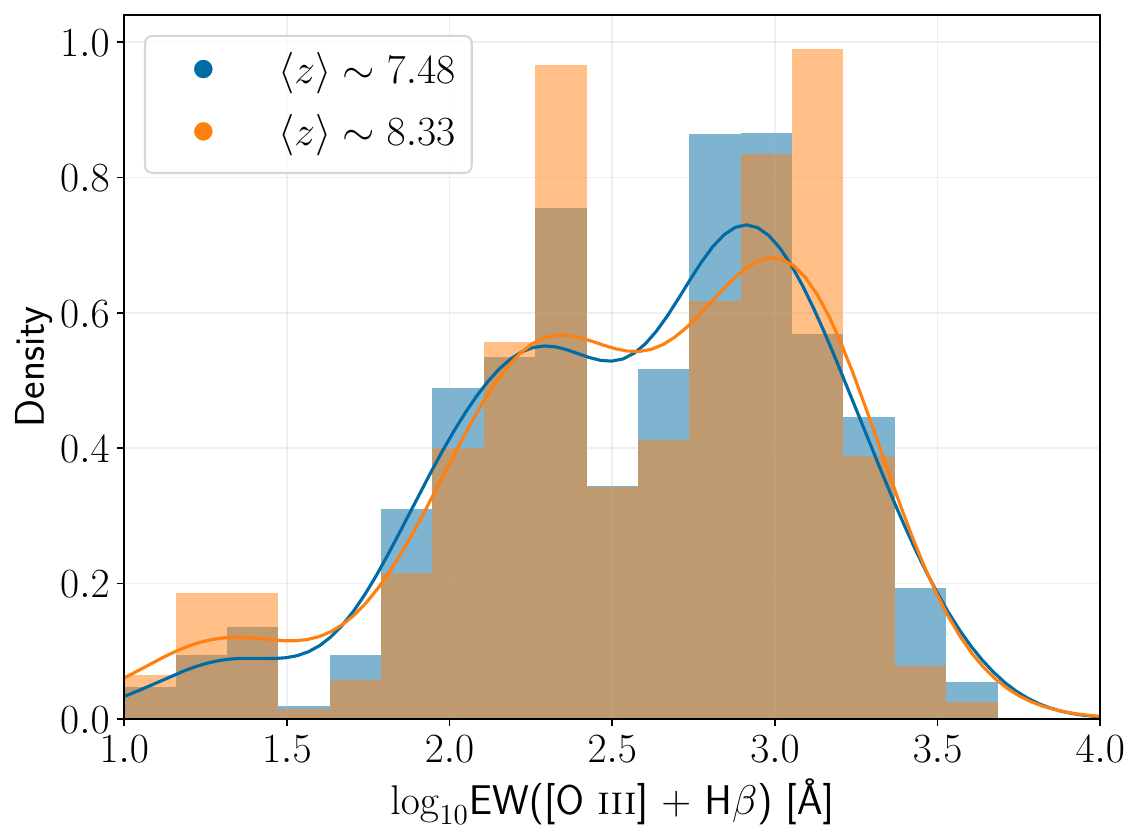}
    \caption{Distribution of the \oiiihb~equivalent width for each redshift bin. This figure is not completeness-corrected. The colours represent the two redshift ranges and the lines are KDE estimate of the distribution for clarity.}
    \label{fig:EW_dist}
\end{figure}

The unique intrinsic depth of GLIMPSE in combination with the strong lensing magnification enables us to derive for the first time the faint-end slope of the \oiiihb~LF (i.e. $\leq10^{41}$ \ergs).
We do find a significant differences between $\alpha$ at redshift $7<z<8$ and $8<z<9$ as we obtain $\alpha = -1.78_{-0.06}^{+0.06}$ and $\alpha = -1.55_{-0.11}^{+0.11}$ respectively, which is slightly above the 1$\sigma$ threshold (see Table~\ref{tab:schechter}). This could be caused by the increasing difficulty of observing the faintest galaxies at higher-redshift, but this effect should be accounted for by completeness correction. However, an evolution of the metallicity could also be an explanation, with higher-redshift galaxies having less time in their history to synthesise metals.\\ 

We can compare this result to \citep{wold_uncovering_2025}, who constrained the faint end of the \Oiiib~LF at $z \sim 7$ from photometric data of the HFF A2744 strongly lensed field, assuming the $\Rthree=6.72$ ratio measured by \citep{sun_first_2023} to infer \Oiiib~from \oiiihb. They obtain a slope $\alpha=-2.07_{-0.23}^{+0.22}$, which is steeper than our result. We investigate possible reasons for this difference here below. 

Firstly, the parametrisation is different as they use a double power law with a fixed value of $\log_{10}L^* = 42$ \ergs, which could cause some different constraints on the fit. However, by refitting their observed LF with a Schechter function fixed at our $\log_{10}L^* \sim 43.0 \pm 0.3$ \ergs~and $\mean{z}\sim7.9$ data from \citet{meyer_jwst_2024} at the bright end, we obtain a similar $\alpha = -2.18_{-0.13}^{+0.13}$. The faint-end slope is compatible between the two parametrisations, so the use of double power law with fixed characteristic luminosity does not explain the difference.

Another difference could be the ranges of faint-end luminosities probed by \citet{wold_uncovering_2025} and GLIMPSE. The former is only constrained by a few data points, with limited detections, down to $L\sim10^{41}$ \ergs, while GLIMPSE reaches to $L\sim10^{39}$ \ergs. 
By fitting our GLIMPSE LF again with a faint-end luminosity limit of $L \geq 10^{41}$ \ergs, which corresponds to the \citet{wold_uncovering_2025} limit, we obtain $\alpha = -2.19_{-0.12}^{+0.16}$ for $\mean{z}\sim7.48$ and $\alpha = -1.92_{-0.25}^{+0.24}$ for $\mean{z}\sim8.33$, which is significantly steeper than our main result, and compatible with \citep{wold_uncovering_2025}. Further details on this can be found in Appendix~\ref{app:LFs}.

\begin{figure}[t]
    \centering
    \includegraphics[width=0.99\linewidth]{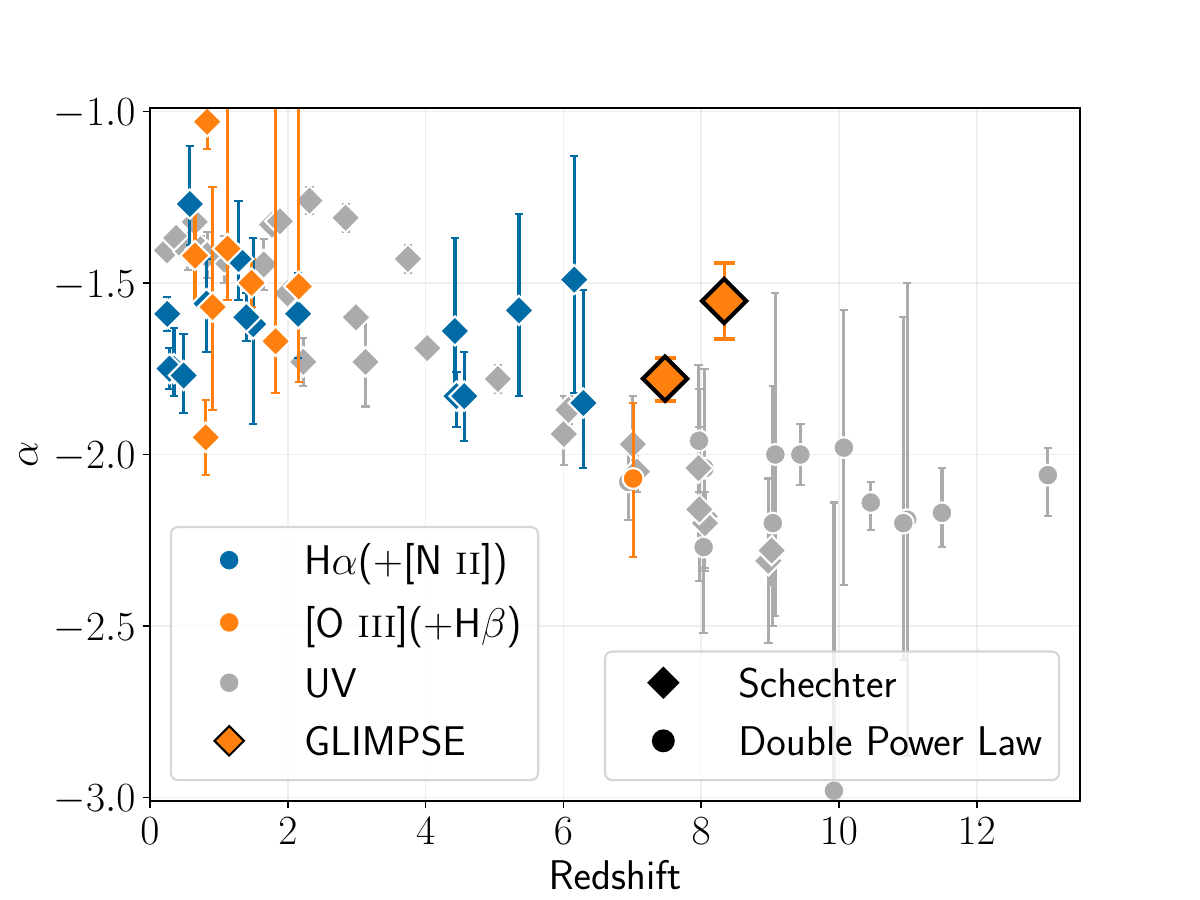}
    \caption{Redshift evolution of the faint-end slope ($\alpha$) in the Schechter (diamond) and double power-law (circle) fits for the UV continuum (grey symbols) and nebular emission line (coloured points) LFs. The orange squares with a black outline correspond to this study. The compilation of the data, including GLIMPSE, can be found in Table~\ref{tab:alpha}.}
    \label{fig:alpha_evo}
\end{figure}

In addition to the bias induced by the differences in depth, the galaxies selected in \citep{wold_uncovering_2025} do not include faint \oiiihb~galaxies. While we do not set limits on the \oiiihb~equivalent width, \citet{wold_uncovering_2025} uses a colour-colour selection comparable to a cut of $\text{EW}(\Oiiib) > 500Å$ (equivalent to $\text{EW}(\oiiihb) > 742Å$ assuming their $\Rthree=6.72$ and $\Oiiib/\Oiiit=2.98$). Figure~\ref{fig:EW_dist} shows the distribution of \oiiihb~equivalent widths for the different redshift bins in GLIMPSE. A significant number of GLIMPSE galaxies ($\sim64\%$) have $\log(\text{EW}_{\oiiihb}) \leq 2.87$, indicating that a significant number of galaxies reside below the detection threshold of \citet{wold_uncovering_2025}, which might skew their statistics \citep{endsley_burstiness_2025, endsley_starforming_2024}. Therefore, the differences between the selection method and survey depth explains the strong difference between the faint end of the GLIMPSE and \citet{wold_uncovering_2025} \oiiihb~LF.\\

\subsection{The faint-end slope of the nebular emission lines and UV luminosity functions}
In Fig.~\ref{fig:alpha_evo} we report the evolution of the faint-end slope $\alpha$ for multiple LFs from past studies at various redshifts. We considered two types of studies: Nebular emission line LFs (\oiii, \ha, \hb) and UV LFs. 
First, we note a clear difference between their redshift evolution. While the faint-end slope of the nebular emission line LF, $\alpha_\text{neb}$, stays approximately constant at $\alpha_\text{neb} \sim -1.62\pm0.21$ (within large uncertainties) from redshift 0 to 9, the faint-end slope of the UV LFs, $\alpha_\text{UV}$, steepens from $\alpha_\text{UV}\sim-1.4$ at $z=0$ to $\alpha_\text{UV}\sim-2.2$ at redshift 9. This means that at high redshift, UV-faint galaxies outnumber UV-bright galaxies, while the proportion of nebular bright to faint galaxies stays approximately constant over time.
This result is independent of the exact parametrisation of the LF --- i.e. the Schechter function \citep{schechter_analytic_1976} or double power law \citep{dunlop_redshift_1990} --- as also shown in this figure.

Several processes or physical causes could explain the redshift evolution difference between the UV and the nebular faint-end slopes:
{\em 1)} bursty or variable star formation histories, leading to an incoherence between \muv~and nebular lines due to their different timescales
{\em 2)} a strong luminosity dependence of the metallicity, causing lower \oiiihb\ emission at fainter luminosities, or
{\em 3)} an intrinsic turnover of the UV LF at the faint-end, but below the range currently observed, possibly flattening the faint end of the \oiiihb~LF. 
We now discuss their effects on the LF one by one.

\begin{figure*}[t]
    \centering
    \includegraphics[width=0.99\linewidth]{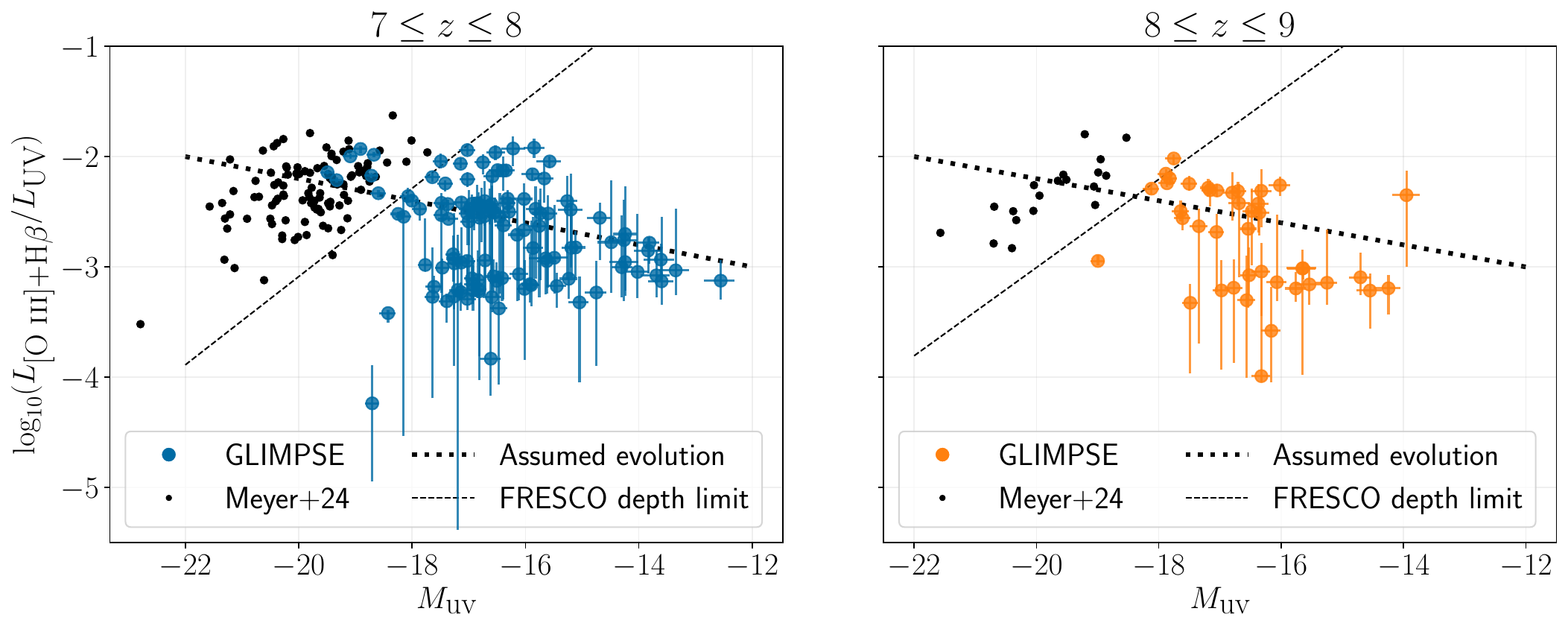}
    \caption{Evolution of the \oiiihb-to-UV ratio with \muv~for the $7<z<8$ sample (left panel) and the $8<z<9$ sample (right panel). In black is the FRESCO data from \citep{meyer_jwst_2024} with depth limit in dashed line, to cover the bright end. The dotted line shows the assumed evolution of the ratio following $R = -0.1\times\muv-4.2$, which enables us to flatten an nebular emission line LF from an intrinsically steep LF, as shown in Fig.~\ref{fig:flattening_O3hb2UV}. This law is arbitrary and chosen to follow the data in a simple manner. However, it does follow closely the linear fit of the data.}
    \label{fig:UV_neb}
\end{figure*}

\subsubsection{The impact of bursty star formation histories on the \oiiihb~LF}

The different timescales of UV continuum or nebular line emission \citep[$\lesssim100$Myr and $\lesssim10$Myr, ][]{kennicutt_star_2012} powered by star formation might cause a decoupling of the slope of the UV and \oiiihb\ LF, as we would expect fainter galaxies to vary on shorter timescales \citep[e.g.][]{endsley_starforming_2024}. Indeed, with variable or bursty star formation histories, these two quantities are expected to evolve differently, resulting in a large scatter between the relative \oiiihb\ and UV emission. If $L(\oiiihb)/\luv$ decreases systematically towards fainter galaxies due to an increased fraction of galaxies with decreasing star formation histories, this could thus in principle explain the relative flattening of the nebular LF with respect to the UV one.

\begin{figure}
    \centering
    \includegraphics[width=\linewidth]{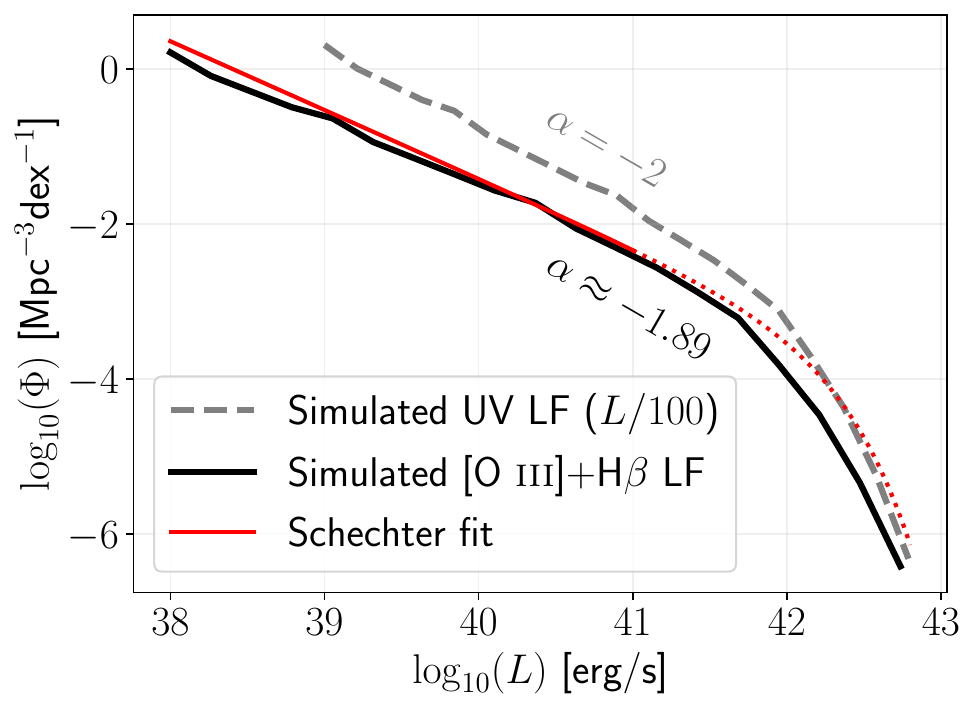}
    \caption{Demonstration of the effects that arise from allowing the nebular emission lines to be fainter than the continuum in the faintest sources. This enables us to flatten an intrinsically steep UV LF ($\alpha\sim-2$, in grey) into a flatter nebular emission line LF ($\alpha\sim-1.89$, in black). In red is the Schechter least square fit ($\alpha=-1.89$, $\log_{10} L^*=42.09$, $\log_{10}\phi_* = -3.63$) of the simulated \oiiihb\ LF. The bright end is dashed, as the effect applied yields non-Schechter-like profile on the bright end.}
    \label{fig:flattening_O3hb2UV}
\end{figure}

In Fig.~\ref{fig:UV_neb}, we show the \oiiihb-to-UV luminosity ratio over \muv~for our two redshift bins. We find that the ratio slowly evolves with \muv, with roughly one order of magnitude change over the whole \muv~range. 
We checked the correlation of the data using a spearman test. For both redshifts bins, the correlation is $>3\sigma$-significant (p-values of $8.14\times10^{-9}$ and $7.07\times10^{-5}$ for $7<z<8$ and $8<z<9$). Then, to test the effect that such an evolution would have on the LF, and following observations from Fig.~\ref{fig:UV_neb}, we consider a log-linear evolution from $\log_{10}(L_{\oiiihb}/L_\text{UV})=$-2 to -3 between $\muv=$-22 to -12 ($R = -0.1\muv-4.2$). By sampling 10 million galaxies from a UV LF with an $\alpha=-2$ and converting them to a \oiiihb~LF using the assumed relation, we obtain a \oiiihb~LF with $\alpha\sim-1.89$ (Fig.~\ref{fig:flattening_O3hb2UV}). This shows that the evolution of the \oiiihb-to-UV ratio has the effect of flattening the \oiiihb~LF, which could explain the difference between our results and some of the observed UV LF. However, most of the UV LFs are steeper than $\alpha=-2$ in Fig.~\ref{fig:alpha_evo} and require more than a bursty star formation.

\subsubsection{The impact of metallicity on the \oiiihb~LF}
\label{sec:impact_Z}

Since \hb~and other H recombination lines are directly related to the number of ionising photons absorbed by the neutral hydrogen (therefore, to the SFR), but forbidden metal lines such as \oiii~also strongly depend on other physical parameters (metallicity and ionisation parameter in particular), it is important to separate their relative contribution to the measurement of \oiiihb~LF reported here. Ideally, this would allow us to infer the pure \hb~LF, providing a cleaner quantity to compare with the UV LF, which, like \hb~also samples massive stars (although on longer timescales).
To achieve this, we now examine the impact of systematic variations of \Oiiib/\hb~as a function of galaxy luminosity.

Indeed, it is well known that $R3 = \Oiiib/\hb$ is on average a bimodal function of metallicity which peaks at $12+\log(O/H)\sim 8$, with a maximum $R3 \sim 6$ (or $(\oiiifull)/\hb \sim 8$), and decreases at higher and lower metallicities \citep[e.g.][]{curti_new_2017, maiolino_re_2019, nakajima_empress_2022, sanders_direct_2024, scholte_jwst_2025}. 
The relation between \muv~and metallicity is complex, but emerges from the mass-metallicity relation, which states that lower-mass galaxies have lower metallicity \citep[e.g.][]{chemerynska_extreme_2024, curti_mass-metallicity_2020, curti_jades_2024, nakajima_empress_2024}. We therefore expect the metallicity to decrease to fainter $M_\text{UV}$ \citep{tremonti_origin_2004, laseter_efficient_2025}. Unfortunately, direct measurements for galaxies as faint as those in our sample require deep medium-resolution and high signal-to-noise spectroscopy. \citet{chemerynska_extreme_2024} studied the dependency of $R3$ on stellar mass for very faint galaxies at redshift $z\sim6-7$, by measuring the metallicities of 8 galaxies to be  $12+\log({\rm O/H})\sim6.70-7.56$ with $\Rthree\sim 1.4-5.5$ at $\muv\sim-15.34$ to $-17.17$. The evolution of  $12+\log({\rm O/H})$ follows the expected decrease in metallicity with \muv~and stellar mass.
The \citet{chemerynska_extreme_2024} sample required  ultra deep spectroscopy and is already very challenging to observe. However, GLIMPSE goes deeper than $M_\text{UV}>-15$ with photometry, which is vastly unexplored at this redshift and might yield lower $\Rthree$, as shown by detections or tentative detections of galaxies with $\Rthree<1$ and \muv~ranging from $-11$ to $-16$ \citep{vanzella_extremely_2023, fujimoto_glimpse_2025, morishita_pristine_2025, hsiao_sapphires_2025}. Such galaxies require higher magnifications and difficult to reach exposure time, making it challenging to statistically analyse them. Therefore, deep photometric studies such as GLIMPSE have their limitation, but enables indirect probes of the effect of metallicity on very faint ($M_\text{UV}>-15$) galaxies. Regarding the brighter end of \oiiihb~fluxes, studies such as \citet{meyer_jwst_2024} measured an extreme $\Rthree=6.38\pm0.85$ line ratio and $\muv\sim -19.65$ for their median stack, close to the maximum observed by \citet{curti_new_2017, maiolino_re_2019}. Such values were also confirmed by other JWST surveys \citep{nakajima_empress_2022, sanders_direct_2024, scholte_jwst_2025}.\\

To account for systematic variations of the \oiii-to-\hb~ratio with \muv, we define the following simple function: 
\begin{equation}
    \Rthreefunc = 
    \begin{cases}
        6.34 & \text{if } \muv < -19\\
        \frac{-1.8 \times \muv - 25.7}{1.34} & \text{if } \muv\in [-19, -16.5]\\
        \frac{-0.5 \times \muv - 4.25}{1.34} & \text{if } \muv\in [-16.5, -12.5] \\
        1.49 & \text{if } \muv \geq -12.5
        
    \end{cases}
    ,
    \label{eq:varying_oiii_hb_ratio}
\end{equation}

\noindent where \Rthreefunc varies between $1.49$ and $6.34$ for $\muv=-12.5$ to $-19$. We used the measurements of \citet{chemerynska_extreme_2024} and \citet{meyer_jwst_2024} as the foundation for this function, and we extrapolated for fainter galaxies to account for our measurements. 
We converted our ratio into the more common \Rthree ratio following  $(\oiiifull)/\hb = 1.34 \times \Oiiib/\hb$. 
Therefore, at the bright end ($\muv\sim-19$), the high \Rthree increases the contribution of \Oiiib~($\sim66\%$) to the total \oiiihb~relative to \hb~($\sim10\%$). However, at the faint end ($\muv\sim-12$), the \Rthree ratio becomes so small that \hb~($\sim33\%$) contributes comparably to \oiiihb~as \Oiiib~($\sim50\%$). This evolving \Rthree ratio with \muv~predicts a steeper \hb~LF and flatter \oiii~LF once \hb~and \Oiiib~are separated. By applying this separation to our LF from Fig.~\ref{fig:O3LF}, we can split the contribution of \hb~and \Oiiib~as shown in Appendix \ref{app:o3lf}. These LFs confirm our assumptions and are parametrised by $\alpha_{\hb} = -1.95_{-0.08}^{+0.08}$ and $-1.68_{-0.14}^{+0.13}$, and $\alpha_{\Oiiib} = -1.66_{-0.05}^{+0.05}$ and $-1.45_{-0.10}^{+0.09}$ for redshift bins $\mean{z}\sim7.56$ and $\mean{z}\sim8.45$, as listed in Table~\ref{tab:schechter}. In short, we now obtain an \hb~LF whose faint-end slope is closer to, but still flatter than, that of the UV LFs at the same redshift ($\alpha_\text{UV} \sim -2\text{ to }-2.2$ at $z\sim8$).

\begin{figure}
    \centering
    \includegraphics[width=\linewidth]{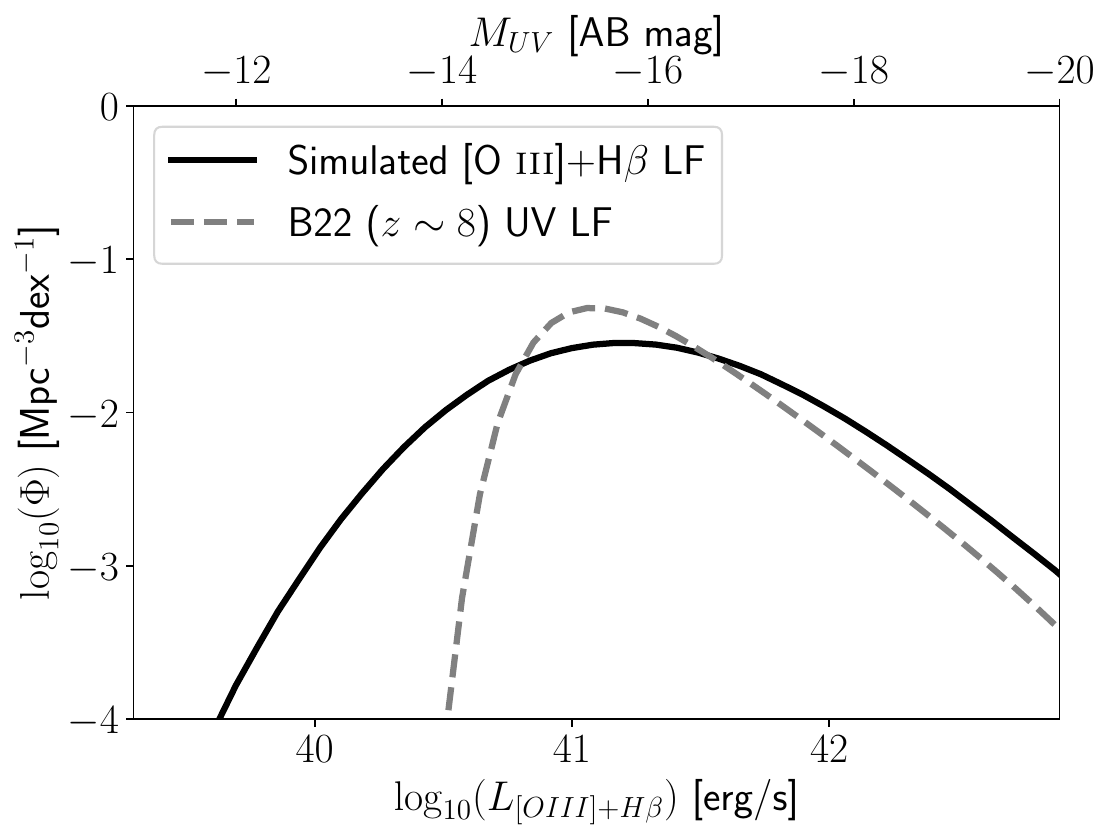}
    \caption{Simulation of the effect of scattering on a sharp faint-end turnover in a Schechter LF. We used the \citep[B22, ][]{bouwens_z_2022} Schechter parametrisation at $z\sim8$, with a sharper turnover $\delta$: $L^* = 10^{45.26}\text{\ergs} ($\muv=-20.9$),~\alpha=-2.2,~\delta=2.0,~L_T=10^{42.98}\text{\ergs}$ ($\muv=-15.2$). 
    We include scatter that correspond to the natural scatter between \oiiihb\ and \muv\ (as seen in Fig.\ \ref{fig:muv_flux}) by sampling from a normal distribution with parameters $\mathcal{N}(\sigma=0.4)$.
    }
    \label{fig:scatter_2_flatten_LF}
\end{figure}

\subsubsection{The early signs of an UV LF turnover}
\label{sec:turnover}

Finally, we explore the third possibility, namely that the flatter \oiiihb\ LF (compared to the UV LF) could result from an intrinsic turnover of the galaxy UV LF at some faint limit $\muv\geq-15.5$ \citep{bouwens_z_2022}.

Previous studies investigated the possibility of finding turnovers in LFs to constrain the formation of the lowest mass haloes. On the theory side, a faint-end turnover of the UV LF is expected as a direct consequence of the photoevaporation of small dark matter haloes during the re-ionisation \citep{shapiro_photoevaporation_2004}. Simulations observed such turnovers \citep[e.g.][]{kuhlen_dwarf_2013, ocvirk_cosmic_2016} by parametrising LFs with an additional turnover parameter \citep[e.g.][]{jaacks_impact_2013} and tentative observations of that turnover were later attempted \citep{bouwens_z_2017, bouwens_z_2022} on the HFF. This tentative detection of a turnover ended up ruling out a turnover in the UV LF above $M_\text{UV} \leq -15.5$ for their $z=2-9$ sample. In more details, at $z\sim6$, they rule out a turnover above $\muv\leq -14.3$. At higher redshift ($z\sim9$), the small number of detected sources back then limited the depth of the LF. Nevertheless, they argued that a turnover above $M_\text{UV} \leq -16$ can be ruled out, but they would be consistent with a turnover at $M_\text{UV}\sim-15$ \citep{bouwens_z_2022, atek_extreme_2018}. According to Fig.~\ref{fig:muv_flux}, this magnitude would roughly translate into a line flux of $L_\oiiihb\sim10^{40}$ \ergs, and therefore, as GLIMPSE observes fainter galaxies, a turnover might be observed. This is roughly compatible with our LFs in Fig.~\ref{fig:O3LF} flattening, but does not prove the existence of this turnover. However, we can simulate the effect of a sharp UV LF faint-end turnover on the \oiiihb~LF. We define this sharp turnover by adding a simple component to Eq.~\ref{eq:schechter}. The final LF is given by 

\begin{equation}
    \phi(L)dL = \phi^*\left(\frac{L}{L^*}\right)^\alpha\exp\left(-\frac{L}{L^*}\right)\exp\left(-\left(\frac{L_T}{L}\right)^\delta\right)d\left(\frac{L}{L^*}\right),
    \label{eq:schechter_turnover}
\end{equation}

\noindent where $L_T$ is the turnover luminosity and $\delta$ turnover parameter for which negative values indicate the presence of a turnover. 
To simulate the scattering effect between the \oiiihb\ and \muv\ (as seen in Fig.~\ref{fig:muv_flux}), we converted $L_{\rm UV}$ to $L_{\oiiihb}$ using the following equation:
\begin{equation}
    \log_{10} L_{\oiiihb} = \log_{10} L_{\rm UV} - 2 + \mathcal{N}(0, 0.4)
    \label{eq:sim_uv_to_oiiihb}
,\end{equation}
\noindent where $\mathcal{N}(0.4)$ is a random number sampled from a normal distribution $\mathcal{N}(\sigma)$, with an additional a factor $\log_{10}(0.01) = -2$ to convert from UV to \oiiihb, following the approximate ratio from Eq.\ \eqref{fig:flattening_O3hb2UV}. The choice of $\sigma$ corresponds to the full-width half-maximum (FWHM) of $\sim 1$dex from Fig.\ \ref{fig:muv_flux}. To show the impact of standard scattering from uncertainties, we sampled a UV LF with arbitrary parameters (as defined in Eq. \eqref{eq:schechter_turnover}) and sampled the \oiiihb\ emission associated with every UV measurement from Eq.\ \eqref{eq:sim_uv_to_oiiihb}. We observe that, while the UV shows a sharp faint-end turnover, the \oiiihb\ LF instead shows a flattening compared to the UV, which can also mimic the observed flattening in Fig.~\ref{fig:O3LF}. 

We also note that because of our very simple scattering assumptions, the bright end is also strongly affected. In reality, the uncertainties are greater for faint galaxies than for brighter ones, and we therefore do not expect to see a major difference in the bright end.
However, with the depth of GLIMPSE, we do not observe such a sharp turnover. This could be biased by the completeness estimation, as the very faint end is very incomplete, which may dominate the LF estimation. This will be discussed in a subsequent GLIMPSE publication (Atek et al. in prep). Finally, UV LFs from JWST surveys did not suggest any turnover in UV at these redshifts or higher, and neither at these relatively similar \muv~\citep[][Atek et al. in prep]{chemerynska_first_2026}. We therefore conclude that the data is not compatible with a turnover in the faint end of the LF.\\

\subsection{Possible missing sources with our method}
\label{sec:missing}
In this last part, we discuss the validity of the completeness correction, in particular the use of the \muv\ completeness correction for our sample. As shown by Fig.\ \ref{fig:muv_flux}, at a fixed \muv, galaxies show variations up to $\sim$1dex in \oiiihb\ flux, meaning that in theory, we might be missing sources by only correcting for completeness on \muv. In essence, this question requires the exploration of the whole range of \oiiihb\ equivalent widths.\\

We first focus on weak \oiiihb\ equivalent widths (e.g. $\lesssim 200-300$Å). Such emission lines are typically difficult to constrain with broadband photometry, as their effect is hard to distinguish from the continuum and the noise, hence these galaxies are solely detected via their continuum. Therefore, with our LBG selection, weak equivalent width galaxies are bounded to the detection of the continuum and should be correctly accounted for by our completeness estimations.

For stronger \oiiihb\ equivalent widths (e.g. $\gtrsim 1000$Å), their effect is visible and significant in the photometry \citep[e.g.][]{smit_evidence_2014}. This increases the signal-to-noise of the affected filters and, in theory, increases the detection rate of galaxies. With our LBG selection, these galaxies would be detected, provided that their continuum is strong enough in other filters (see Eq. \eqref{eq:selection}). However, there might be situations where the continuum is not well enough detected (S/N$_{\rm F115W}<5$ and S/N$_{\rm F200W}<5$; see Eq. \eqref{eq:selection}), but the filters affected by \oiiihb\ are still detected thanks to their strong emissions. In that case, the LBG selection does not find these sources, but are in fact not accounted for in our final completeness estimation. Correcting for this effect would require additional exploration of the completeness, by injecting galaxies with realistic \oiiihb\ at a given \muv.

The current statistics of $z=7-9$ galaxies with ultra-faint \muv\ is relatively low, making galaxy templates exploratory in this regime. But in addition to that, such UV-faint and strong \oiiihb\ galaxies might simply be rare in the Universe. \citet{endsley_starforming_2024} observed a decrease in the average \oiiihb\ equivalent width with fainter galaxies, with almost one dex difference between $\muv\sim-20.1\ {\rm and}\ -17.6$ at $z=7-9$. With our sample going down to $\muv \sim -11$, the decline might continue and making these galaxies even rarer. Therefore, the impact of these galaxies on the final LF should be limited.

\section{Implications and discussion}
\label{sec:implication}
We now present and discuss the main implications of our nebular LFs. We first use them to estimate the ionising photon budget produced by our galaxies, and then we infer the cosmic star formation rate at high redshift ($z \sim 7-9$), where measurements of \ha\ are not feasible anymore with NIRCam. All of the following will assume the Eq.~\ref{eq:varying_oiii_hb_ratio} function for the $\Rthree$ ratio.

\subsection{The ionising photon budget at $z \sim 7-9$}
\label{sec:impl_ion}
For the Universe to re-ionise at its measured rate \citep{bouwens_reionization_2015, mason_model-independent_2019}, galaxies needs to reach some ionising photon-production budget to overcome for recombination. The source of ionising photons comes from multiple objects \citep[Star-forming galaxies and AGNs][]{robertson_cosmic_2015, robertson_new_2013}. Focusing on star formation, Balmer series emission lines enable us to estimate the ionising photon-production rate from star-forming galaxies.
From the \oiiihb~LF, we can infer the \hb\ (or \ha\ LF), which can then be summed to obtain the instantaneous ionising photon-production rate of galaxies, and thus, for a known escape fraction of ionising photons $f_\text{esc}$, the ionising photon emissivity $\dot{N}_\text{ion}$ from galaxies. It can be measured using the following equation:

\begin{align}
    \dot{N}_\text{ion} &= \int_{L_\text{min}}^\infty \fesc Q_\text{ion}(L)\phi(L) dL\\
    &= \int_{L_\text{min}}^\infty \fesc\frac{L}{c_\alpha (1-\fesc)}\phi(L) dL,
    \label{eq:emissivity}
\end{align}

\noindent where $Q_\text{ion}(L)$ is the ionising photon-production rate, which is related to the luminosity $L = Q_\text{ion}c_\alpha(1-f_\text{esc})$, $c_\alpha$ is the ratio between the line emissivity and the total recombination rate for which we adopt $c_\alpha = 1.37\times10^{-12}$ erg for an electron temperature of $10^4$ K \citep{schaerer_transition_2003}, $\fesc$ is the escape fraction of ionising photons and $\phi(L)$ is the LF.  
In practice, we convert the measured \oiiihb\ flux to \hb\ using the ratio given by Eq.~\ref{eq:varying_oiii_hb_ratio}, we correct \hb\ for dust attenuation (see Sect.~\ref{sec:dust_att}) and then we multiply by 2.86 \citep{osterbrock_astrophysics_2006} to obtain the dust-corrected \ha\ LF shown in Fig.~\ref{fig:HaLF}. 
To constrain the bright end, we converted the data from \cite{meyer_jwst_2024} in the same manner. However, as they reported a negligible dust attenuation in their stacked spectra, we directly converted their \Oiiib~data using their stacked ratio $R3\sim 6.38 \pm 0.85$. 
As expected, the correction for dust is overall rather small, and the resulting LFs are therefore similar
(see Figs.~\ref{fig:HbLF} and \ref{fig:HaLF}), as also seen from Table~\ref{tab:schechter}.\\

\begin{figure*}[t]
    \centering
    \includegraphics[width=\linewidth]{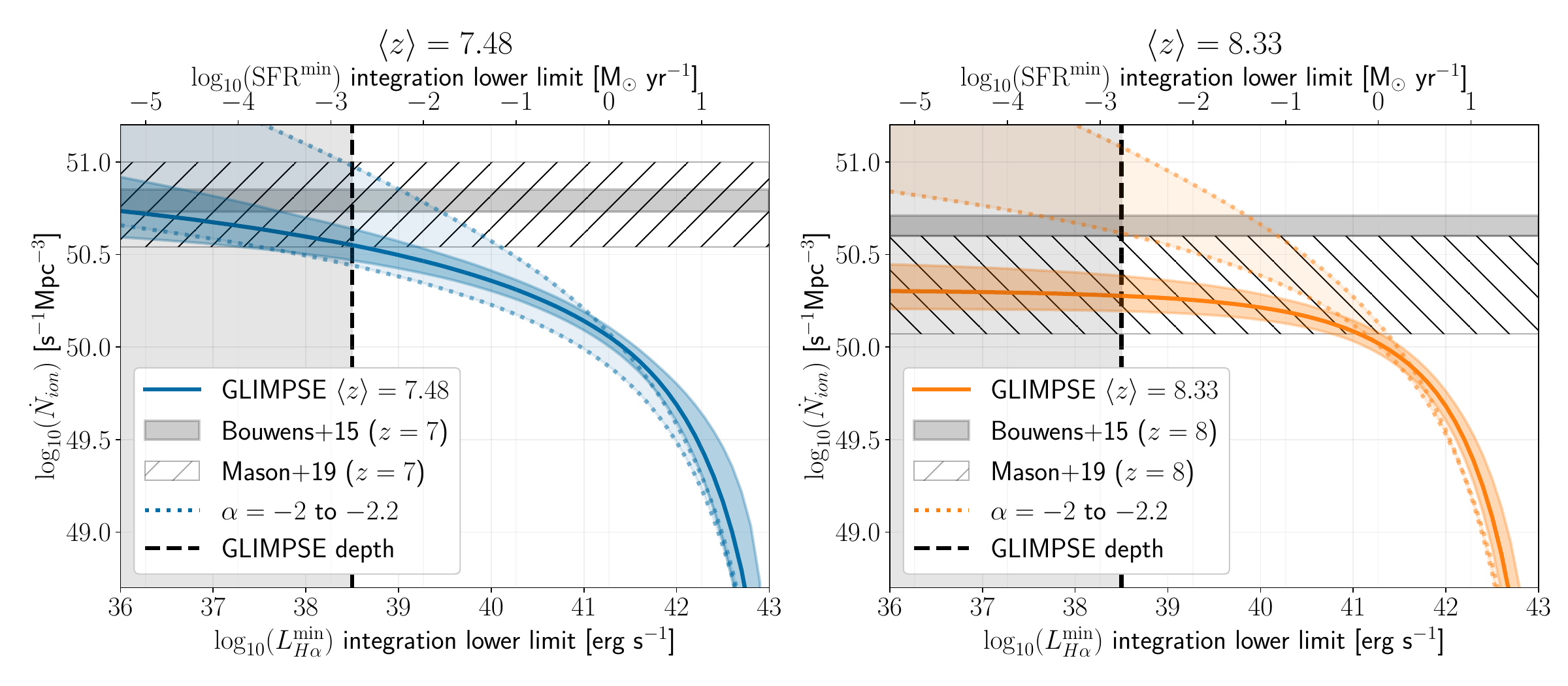}
    \caption{Ionising photon emissivity as a function of the minimum H$\alpha$ luminosity $L_{\ha}^{\rm min}$ as extrapolated from our observed \oiiihb~LF (see Sect. \ref{sec:impl_ion}) in the two redshift bins we investigated. 
    The ionising photon emissivity required to re-ionise the Universe at its measured rate, taken from \citet{bouwens_reionization_2015} is shown by the shaded areas and from \citet{mason_model-independent_2019} in the hatched areas. All of our measurements assume $\fesc=0.14$ \citep{jecmen_glimpse_2026}, the solid lines are extrapolated from this work and the dotted lines assumes $\alpha=-2\text{ to }-2.2$, which is similar to the measured faint-end slope of the typical UV LF at this redshift. The black line shows the GLIMPSE depth following the data from Fig.~\ref{fig:HaLF}.}
    \label{fig:ionisation_budget}
\end{figure*}

Figure~\ref{fig:ionisation_budget} shows the resulting ionising photon emissivity at $z\sim7-9$ from GLIMPSE, as a function of the lower integration limit $L^{\rm min}_{\ha}$, and for a standard fixed $\fesc=0.14$ scenario \citep{jecmen_glimpse_2026}.
Probably the most striking result is that, since our faint-end slope is $\alpha > -2$, the cumulative emissivity becomes already fairly flat over the range of \ha\ luminosities probed by GLIMPSE. For example, for an integration lower limit $L_{\ha} = 10^{39}$ \ergs, the ionising photon emissivity reaches $\log_{10}\dot{N}_{\rm ion} = 50.50_{-0.07}^{+0.07}$ and $\log_{10}\dot{N}_{\rm ion} = 50.26_{-0.08}^{+0.10}$ for our two redshift bins ($7<z<8$ and $8<z<9$), which corresponds to 44\%-58\% and 36\%-46\% of the total re-ionisation ionising photon budget measured by \citet{bouwens_reionization_2015}. However, for the more recent estimations of $\dot{N}_{\rm ion}$ by \citet{mason_model-independent_2019}, our galaxies account for $31\%-90\%$ and $46\%-156\%$ of the budget for our two redshift bins ($7<z<8$ and $8<z<9$), implying that star-forming galaxies provide sufficient ionising photons to re-ionise the Universe.
Adopting a lower integration limit, e.g.~$L_{\ha} \sim 10^{38}$ \ergs, increases the total contribution of ionising photons only by about $1\%-10\%$. This means that, for a scenario with constant $f_{esc}$, the  population of galaxies with very low star formation rates SFR$\la (0.005-0.001)$ \msunyr, below the detection limit of GLIMPSE, contributes only a small fraction of the total production of ionising photons. The inferred values of \nion\ exhibit variations depending on the assumed constant escape fraction, \fesc, since \nion\ scales with $\frac{\fesc}{1-\fesc}$. Specifically, \nion\ is 0.5dex lower for $\fesc=5\%$ or 0.2dex higher for $\fesc=20\%$.\\

According to our earlier scenario, the flattening of the \oiiihb~LF is driven by a population of faint galaxies characterised by bursty star formation history and lower metallicities. 
In this picture, many of these faint galaxies are not actively forming stars, thus diminishing their contribution to the ionising photon budget for re-ionisation. Consequently, the inferred \nion\ stabilises at a fixed value.

To increase the contribution of fainter galaxies, scenarios involving varying escape fraction \fesc enables more ionising photons to escape and thus enhancing their role in cosmic re-ionisation. Such scenarios could emerge from the variation of \fesc with galaxy mass \citep[e.g.][]{naidu_synchrony_2022, begley_vandels_2022, flury_low-redshift_2022, saldana-lopez_vandels_2023, pahl_connection_2023}, metallicity \citep[e.g. for dwarf irregulars][]{ramambason_inferring_2022, hunter_interstellar_2024}, SFR \citep[e.g.][]{giovinazzo_breaking_2025}, $\beta$-slope \citep[e.g.][]{chisholm_far-ultraviolet_2022, giovinazzo_breaking_2025} or else. We refer to the GLIMPSE paper by \citet{jecmen_glimpse_2026}, which studies the \fesc~in more detail.

For comparison, we also display in Fig.~\ref{fig:ionisation_budget} the ionising photon emissivity inferred from the \ha~LF, adopting $\alpha\sim-2\text{ to }-2.2$ (And the same $\phi^*$ and $L^*$ as the \ha\ LF). This slope corresponds to the range measured for some UV LF (see Sect.~\ref{sec:faint_end}). Due to the shape of the Schechter function, when the faint-end slope $\alpha$ is smaller than $-2$, the ionising photon emissivity does not converge; instead, it continues to increase. 
This leads to a rapid overshooting of the ionising photon budget due to the enhanced contribution of fainter galaxies in the cosmic re-ionisation in that scenario. This problem has earlier been referred to as the ionising photon crisis, which arise by measuring the ionising photon-production rate from the UV LF. This requires the knowledge of the ionising photon-production efficiency \xion. By assuming a high value for \xion\ and an increase in \fesc\ for UV-faint galaxies, the Universe re-ionises much too rapidly \citep[see][]{munoz_reionization_2024}.

Our finding of a flattening of the LF therefore argues in favour of their third solution, where lower-mass galaxies have a diminishing impact on the ionising photon-production rate, in contrast to scenarios of low-mass galaxy driven re-ionisation proposed by other studies \citep[e.g.][]{simmonds_low-mass_2024, atek_most_2024}. By computing $\xi_{\rm ion} = Q_{\rm ion} / L_{\rm UV}$ from our \ha~and \muv~measurements, we obtain $\log_{10}\xi_{\rm ion} = 25.31_{-0.62}^{+0.45}$ \Hzerg\ for $\mean{z}\sim 7.48$ and $25.31_{-0.59}^{+0.45}$ \Hzerg\ for $\mean{z}\sim 8.33$, using $\fesc = 0.14$. For comparison with other paper, we also provide $\log_{10}\xi_{\rm ion, 0} = \log_{10}\xi_{\rm ion} + \log_{10}(1-\fesc) = 25.24_{-0.61}^{+0.45}$ \Hzerg\ and $25.31_{-0.59}^{+0.45}$ \Hzerg. The measurement does not evolve between our two redshift bins, and shows a decrease with increasing \muv~(see Fig.~\ref{fig:UV_neb}).  We refer the reader to the GLIMPSE paper by Chisholm et al (in prep) for a detailed analysis of $\xi_{\rm ion}$ at $z\sim6$.

This $\xi_{\rm ion}$ value is lower than measurements from the early-JWST $\xi_{\rm ion}$ \citep[e.g.][]{simmonds_low-mass_2024, atek_most_2024}, which measured higher ionising photons production efficiency for fainter galaxies. However, later results from \citet{simmonds_ionizing_2024}, which included lower-mass galaxies with a JWST photometric analysis, lead to a decrease in the ionising photon-production efficiency $\xi_{\rm ion}$, enabling them to reduce the amount of ionising photons produced and not overestimate the ionising photon emissivity $\dot{N}_{\rm ion}$ from \citet{munoz_reionization_2024}. Our findings agrees with this later result and does not overshoot the ionising photon emissivity required to re-ionise the Universe.\\

\subsection{The cosmic star formation rate density}

The intensity of Balmer series lines such as H$\alpha$ are closely related to the SFR of the galaxy \citep{kennicutt_star_2012}. We obtain the instantaneous SFR using the \ha~line flux as follows:

\begin{equation}
    \text{SFR} = \frac{L_{H\alpha}}{C_X}~\text{ [}M_\odot\text{ yr}^{-1}\text{]}
    \label{eq:Ha2SFR},
\end{equation}

\noindent with the constant of proportionality $\log_{10}(C_X) = 41.27$ calibrated by \citet{kennicutt_star_2012}. The calibration was performed with the Kroupa IMF \citep{kroupa_galactic-field_2003}, but because of the similarities between Kroupa and Chabrier IMF, the difference is minimal and therefore, we neglect conversion factors. The final SFR LF has the exact same shape as Fig.~\ref{fig:HaLF}, with a simple horizontal shift due to conversion to SFR. We included the SFR conversion on top of Fig.~\ref{fig:HaLF}.\\

Finally, to deduce the cosmic star formation rate density (SFRD), we simply integrate the \ha~LF between a minimum luminosity value and theoretically infinity using

\begin{equation}
    \text{SFRD} = \int_\text{SFRD$_\text{min}$}^{+\infty} \text{SFR}\times\Phi(\log\text{SFR})d\log\text{SFR}
    \label{eq:sfrd}
.\end{equation}

We numerically integrate the SFR LF between a given lower limit and $10^5$~\msunyr to obtain the SFRD following Eq.~\ref{eq:sfrd}. Because GLIMPSE goes deeper than previous studies, who set a lower integration limit at  ${\rm SFRD}_{\rm min} \sim 0.24-0.30$ \msunyr (hereafter, standard). This corresponds to $\log_{10}(L_{\ha}/\text{\ergs})\sim40.75$, which only covers the brighter galaxies of GLIMPSE and discard the faint-end. Therefore, we will compare our results with the same integration limit, as well as the deeper limit of $0.005$ \msunyr (hereafter: deep, and approximately corresponding to the faint end of the GLIMPSE detections of $L_{\ha}\sim 10^{39}$ \ergs). In addition, while in principle Eq.~\ref{eq:sfrd} must be integrated to infinity, the uncertainties due to unexplored parameter space and numerical errors would artificially increase the measurement uncertainties at the bright end. Therefore, to avoid it, we set a reasonable integration upper limit of $1000$ \msunyr (i.e. $L_{\ha} \sim 10^{44}$ \ergs).

\begin{figure}
    \centering
    \includegraphics[width=0.99\linewidth]{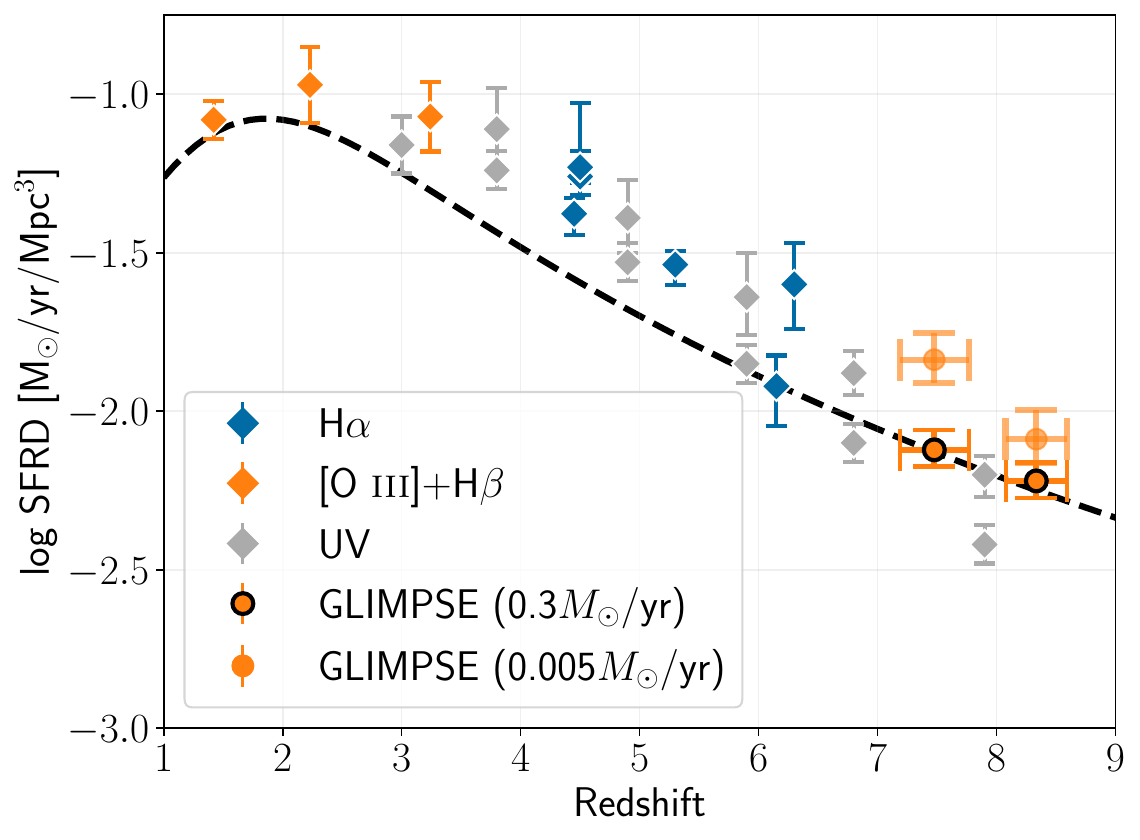}
    \caption{Evolution over redshift of the star formation rate density. We used the same colour scheme as Fig.~\ref{fig:alpha_evo}, where grey corresponds to measurements based on UV, blue to H$\alpha$ and orange to \oiii~and \oiiihb. The black line corresponds to the results of the review \cite{madau_cosmic_2014}, which compiles the state-of-the-art measurements of the time. All the results have an integration lower limit of $0.24-0.30$ \msunyrvol and are IMF-corrected to \citet{chabrier_galactic_2003}. The $z\sim7-9$ orange points correspond to GLIMPSE measurements, where the bright orange uses the standard 0.3\msunyrvol\ lower integration limit and the light orange uses the deep 0.005\msunyr\ lower integration limit (approximately corresponding to $L_{\ha}\sim10^{39}$ \ergs). The non-exhaustive compilation of UV, \ha~and \oiiihb~based measurements, including this work, are listed in Table~\ref{tab:SFRD}.}
    \label{fig:SFRD}
\end{figure}

In Fig.~\ref{fig:SFRD} we observe the redshift evolution of the cosmic star formation rate density (SFRD). We showed multiple literature points, coming from both UV measurements \citep{oesch_dearth_2018, bouwens_alma_2020} and \ha~measurements \citep{bollo_h_2023, covelo-paz_h_2025, fu_medium-band_2025}, as well as two sets of GLIMPSE data points with different integration limits. The bright orange data point corresponds to standard integration limit, while the light orange to a deep integration limit of $0.005$ \msunyr. 
{GLIMPSE reaches deeper line fluxes, and shows flattening that suggest that fainter galaxies do not significantly contribute to the star formation rate density.}
GLIMPSE reaches deeper line flux limits and reveals a flattening in the LFs, suggesting that the contribution of fainter galaxies to the star formation rate density is not significant.
Therefore, we decided to add a secondary measurement at lower integration limit. Table~\ref{tab:SFRD} gives the SFRD for both integration limits and both redshift ranges, as well as the compilation of literature data points shown in Fig.~\ref{fig:SFRD}.

Our results with the standard integration limits seem to be inline with an extrapolation of the \ha~literature measurement, but it differs from UV measurements. However, both UV measurement comes from pre-JWST surveys, which had limited statistics in their faintest bins. All the nebular emission line measured SFRD from \ha~come from JWST results, which is more sensitive to fainter galaxies, allowing us to retrieve the total SFRD better. In addition, UV and nebular emission lines do not trace the same star formation \citep{kennicutt_star_2012}, which indicates a burstier star formation in the early Universe. Therefore, GLIMPSE does show a slightly higher total SFRD compared to previous UV-focus SFRD surveys, but inline for nebular emission lines and the compilation of \citet{madau_cosmic_2014}.

Between the two integration limits, we observe a mild increase in the SFRD for the deeper integration limit, of 0.3dex ($\sim\times1.99$) and 0.1dex ($\sim\times 1.26$) for $\mean{z}\sim7.48$ and $\mean{z}\sim8.33$ respectfully. This increase is only significant for the $\mean{z}\sim 7.48$ redshift bin, which shows an increasing importance of lower star-forming galaxies to the total SFRD with lowering redshift. This nevertheless agrees with the result showed in the previous section, with the fainter galaxies having a more limited impact on the total SFRD, with an addition of only a fraction of dex of SFRD despite the almost 2 orders of magnitude deeper data.

\section{Conclusions}
We measured the \oiiihb~LF for the strongly lensed field Abell S1063 with very deep JWST GLIMPSE observations. We selected our sample of 164 unique galaxies between redshift $7<z<9$ using an LBG selection in combination with photometric redshifts obtained via SED fitting. We constrained the strongly lensed field using very deep HST and JWST images, removed the multiple counter images of some galaxies, and measured the completeness of our sample. We used an SED fitting to measure the \oiiihb~flux for each galaxy, the dust attenuation, and other physical quantities, and we then constructed the \oiiihb~LF for that sample. Combined with existing measurements at the bright end from \cite{meyer_jwst_2024}, we thus determined the \oiiihb~LF down to unprecedented luminosities, reaching $L_{\oiiihb} \geq 10^{39}$ \ergs (equivalent to galaxies as faint as $\muv\sim-12$; the average completeness at these magnitudes is low, but the bins have multiple detections due to magnification). The main results of our study are summarised below.
\begin{enumerate}
    \item The \oiiihb~LFs show a relatively flat faint-end slope $\alpha \sim -1.55\text{ to }-1.78$ at $z\sim7-9$, which contrasts with previous LFs. Studies such as \citet{wold_uncovering_2025} already paved the way into fainter LFs by combining photometry of Abell 2744 and its lensing power, which enabled them to reach $L_{\oiiihb} \geq 10^{41}$ \ergs. They measured a faint-end slope $\alpha=-2.07_{-0.23}^{+0.22}$, which is significantly steeper than that of GLIMPSE. By fitting the GLIMPSE LF with a similar luminosity limit, we measured faint-end slopes $\alpha$ comparable to their results. On the other hand, spectroscopy-based LFs have more limited access to the faintest galaxies, with the best studies reaching $L_{\oiiihb} \geq 10^{41.75}$ \ergs. \citep[e.g.][]{meyer_jwst_2024, matthee_eiger_2023, sun_first_2023, de_barros_greats_2019}. Our best comparison is with \citet{meyer_jwst_2024}, but their sample typically probed the faint end of the LF only in a limited way, and they fixed the faint-end slope to steep values of $\alpha<-2$.
    \item This result contrasts with the steeper slope observed in UV LFs ($\alpha \lesssim -2$) generally found at $z\ga 7$. To explain these differences between LFs of nebular emission lines and the UV continuum, we examined three  scenarios: 

    {\em i)} A decrease in the \oiiihb-to-UV ratio towards fainter galaxies (Fig.~\ref{fig:muv_flux}), which might be caused by more UV-faint (i.e. lower-mass) galaxies seen in a downturn phase \citep{endsley_burstiness_2025}.
    {\em ii)} A decrease in \Rthree=\Oiiib/\hb~towards fainter galaxies due to the decreasing metallicity, as observed by recent studies \citep[e.g.][]{chemerynska_extreme_2024, meyer_jwst_2024}, which implies that the \hb~and the \Oiiib~LFs have different faint-end slopes (Fig.~\ref{fig:flattening_O3hb2UV}). 
    {\em iii)}  A possible sharp turnover of the UV-LF at the faint end below the current detection limits, which, due to natural scattering, would be observed as a simple flattening of the LF (Fig.~\ref{fig:scatter_2_flatten_LF}). 
    As no flatteing of the UV LF has so far been found, but burstiness and the evolution of $\Rthree=\Oiiib/\hb$ have both been observed, we favour the first two solutions (i and ii) to explain the flatter nebular LF compared to the UV LF.

    \item Assuming an average relation between the \Rthree ratio and \muv\ (Eq.~\ref{eq:varying_oiii_hb_ratio}), we separated the contribution of \hb~and \Oiiib\ to the observed \oiiihb, leading to a steeper \hb~LF ($\alpha \sim -1.68\text{ to }-1.95$; Fig.~\ref{fig:HbLF}) and a flatter \Oiiib~LF ($\alpha \sim -1.45\text{ to }-1.66$; Fig.~\ref{fig:O35007LF}). Therefore, the \hb~LF approaches but remains flatter than the UV LF ($\alpha \leq -2$), while the \Oiiib~LF quickly flattens out.

    \item Correcting for dust attenuation (measured through an SED fitting), we converted the \hb~LF into a dust-corrected \ha~LF (Fig.~\ref{fig:HaLF}), which allowed us to compute the total ionising photon emissivity and the cosmic SFR density of galaxies at $z \sim 7-9$ (Fig.\ \ref{fig:ionisation_budget}). Since the slope $\alpha_{\rm nebular} > -2$, we found that the total photon emissivity quickly saturates and does not significantly increase when it is integrated to objects fainter than those observed with GLIMPSE. Assuming a standard constant $\fesc=0.14$, we reached $31\%-90\%$ and $46\%-156\%$ of the ionising photon budget required to drive re-ionisation at $z \sim 7.5$ and $z \sim 8.3$, respectively, as measured by \cite{mason_model-independent_2019}. 
    \item The SFRD (Fig.\ \ref{fig:SFRD}) showed comparable results to previous studies for comparable lower integration limits ($\sim0.3\msunyr$). Integrating down to our observed limit of ${\rm SFR}\sim 0.005\msunyr$ (corresponding to $L_{\ha}\sim10^{39}$\ergs), we obtained an SFRD higher by $\sim 0.1-0.3$ dex, where the increase is only significant at $\mean{z}\sim7.48$. The inclusion of fainter, lower SFR galaxies will not strongly increase the cosmic SFRD at these redshifts because the observed nebular LF, which provides the most robust measure of the instantaneous SFRD, is significantly flatter than the UV LF. 
\end{enumerate}

In short, the determination of the LF of nebular emission (\oiiihb) from star-forming galaxies down to very faint fluxes yields flatter slopes ($\alpha \sim -1.55\text{ to }-1.78$) of the LF than previously thought and measured for the UV LF at high-$z$ ($z \ga 7$). This indicates that GLIMPSE, combining ultra-deep JWST observations and strong gravitational lensing, has allowed us to reach the bulk of the star-forming galaxies at $z \sim 7-9$, and thus, to determine their total ionising photon emissivity and the total cosmic SFRD of these objects. Our results suggest that faint galaxies contribute less to cosmic re-ionisation than previously thought because their number density flattens too rapidly to maintain a highly ionising photon-production rate. 
With conventional or plausible assumptions on the escape fraction of ionising photons, the observed galaxies are capable of driving cosmic re-ionisation at $z \sim 7-9$.

\begin{acknowledgements}
DK thanks Emma Giovinazzo, Andrea Weibel, Callum Witten, Max Briel, Mengyuan Xiao for their useful discussions and suggestions throughout this project.
This work is based on observations made with the NASA/ESA/CSA \textit{James Webb} Space Telescope. The data were obtained from the Mikulski Archive for Space Telescopes at the Space Telescope Science Institute, which is operated by the Association of Universities for Research in Astronomy, Inc., under NASA contract NAS 5-03127 for JWST. These observations are associated with program $\#03293$.
Support for program $\#03293$ was provided by NASA through a grant from the Space Telescope Science Institute, which is operated by the Association of Universities for Research in Astronomy, Inc., under NASA contract NAS 5-03127.
IC acknowledges funding support from the Initiative Physique des Infinis (IPI), a research training program of the Idex SUPER at Sorbonne Université.
LJF and AZ acknowledge support by Grant No. 2020750 from the United States-Israel Binational Science Foundation (BSF) and Grant No. 2109066 from the United States National Science Foundation (NSF); by the Ministry of Science \& Technology, Israel; and by the Israel Science Foundation Grant No. 864/23.
HA and IC acknowledge support from CNES, focused on the JWST mission and the Programme National Cosmology and Galaxies (PNCG) of CNRS/INSU with INP and IN2P3, co-funded by CEA and CNES. HA is supported by the French National Research Agency (ANR) under grant ANR-21-CE31-0838.
RAM acknowledges support from the Swiss National Science Foundation (SNSF) through project grant 200020\_207349.
ASL acknowledges support from Knut and Alice Wallenberg Foundation.
AA acknowledges support by the Swedish research council Vetenskapsr{\aa}det (VR 2021-05559, and VR consolidator grant 2024-02061).
JBM acknowledges support from NSF Grants AST-2307354 and AST-2408637.
This work has received funding from the Swiss State Secretariat for Education, Research and Innovation (SERI) under contract number MB22.00072, as well as from the Swiss National Science Foundation (SNSF) through project grant 200020\_207349. The Cosmic Dawn Center (DAWN) is funded by the Danish National Research Foundation under grant DNRF140.
Softwares: Emcee \citep{foreman-mackey_emcee_2013}, Astropy \citep{collaboration_astropy_2022}, Numpy \citep{harris_array_2020}, Scipy \citep{virtanen_scipy_2020}, Matplotlib \citep{hunter_matplotlib_2007}, Seaborn \citep{waskom_seaborn_2021}, Photutils \citep{bradley_photutils_2024}, LMFIT \citep{newville_lmfit_2025}.

\end{acknowledgements}

\bibliographystyle{aa}
\bibliography{newrefs.bib}

\appendix

\section{Validating the SED fitting flux and EW measurement using flux excess}
\label{app:validation}
To assess the quality of CIGALE measurements on the GLIMPSE dataset, we measure an estimation of the \oiiihb~line flux and the equivalent width according to the excesses measured using the F444W, F410M and F480M filters. We assumed the line flux of \oiiifull~to be separate from \hb, giving us two unknown quantities, for which we need two equations. The reason for this separation is to account for cases where \hb~is in a filter, and \oiii~is in another. In addition to that, we consider the continuum flux as another unknown quantity, requiring an additional equation. We then considered three equations of the same form, given by Eq. \eqref{eq:flux_eq}: 

\begin{equation}
    W_\text{eff}^\text{A}F_\lambda^\text{A} = W_\text{eff}^\text{A}F_\text{$\lambda$, cont} + N_{\hb}^\text{A} \times F_{\hb} + N_\text{\oiii}^A \times F_\text{\oiii}
    \label{eq:flux_eq}
,\end{equation}where $A$ is the filter of interest, $W_\text{eff}$ is the effective width of the filter in Å, $F_\lambda$ is the spectral flux density in erg/s/cm$^2$/Å, $N_{\hb;\oiii}$ is the coverage coefficient (which describes the proportion of the line covered by a filter compared to peak) and $F_{\hb;\oiii}$ is the monochromatic flux density in erg/s/cm$^2$.

With three filters, we can rewrite this system as Eq. \eqref{eq:flux_matrix} with A, B, and C being the three filters of interest. Because these equations mix monochromatic flux density and spectral flux densities, the vectors and the matrix are not uniform in units.

\begin{equation}
    \begin{pmatrix}
        W_\text{eff}^\text{A}F_\lambda^\text{A}\\
        W_\text{eff}^\text{B}F_\lambda^\text{B}\\
        W_\text{eff}^\text{C}F_\lambda^\text{C}
    \end{pmatrix}
    =
    \begin{pmatrix}
        N_{\hb}^{A} & N_\text{\oiii}^{A} & W_\text{eff}^\text{A}\\
        N_{\hb}^{B} & N_\text{\oiii}^{B} & W_\text{eff}^\text{B}\\
        N_{\hb}^{C} & N_\text{\oiii}^{C} & W_\text{eff}^\text{C}
    \end{pmatrix}
    \begin{pmatrix}
        F_{\hb}\\
        F_\text{\oiii}\\
        F_\text{$\lambda$, cont}
    \end{pmatrix}
    \label{eq:flux_matrix}
\end{equation}

To compute the coverage coefficient $N$, which corresponds to the amount of line flux that we expect to retrieve in each filter for a given redshift, we used the convolution of the filter with a synthetic emission line.
We built the synthetic emission lines separately for \oiiihbfull. In the former, we used a centred Gaussian of standard deviation 40Å, and for the latter two Gaussians separated by physical distances and a standard deviation of 40Å. We assumed a ratio of 2.98 between \oiii$\lambda5007$Å and \oiii$\lambda4960$Å \citep{storey_theoretical_2000}. Next, we convolved these synthetic emission lines with the JWST filters of interest and obtained a value of convolution depending on the wavelength. By normalising the convolution of the line at the redshift of interest by the maximum convolution, we obtain a coefficient $N\in[0, 1]$, which we can measure for each line and filter. In the cases where the lines of interest lie outside medium filters or when medium filters are not detected or available, this method does not hold. When outside of the medium filters, an entire row of the coefficient matrix becomes zeros, leading to a singular matrix. In addition, when the medium filter is not detected, we lack one of the two equations. In these cases, we still try to estimate the \oiiihb~line flux by combining \oiiihb, which reduces the number of equations needed.\\

\begin{figure}
    \centering
    \includegraphics[width=0.8\linewidth]{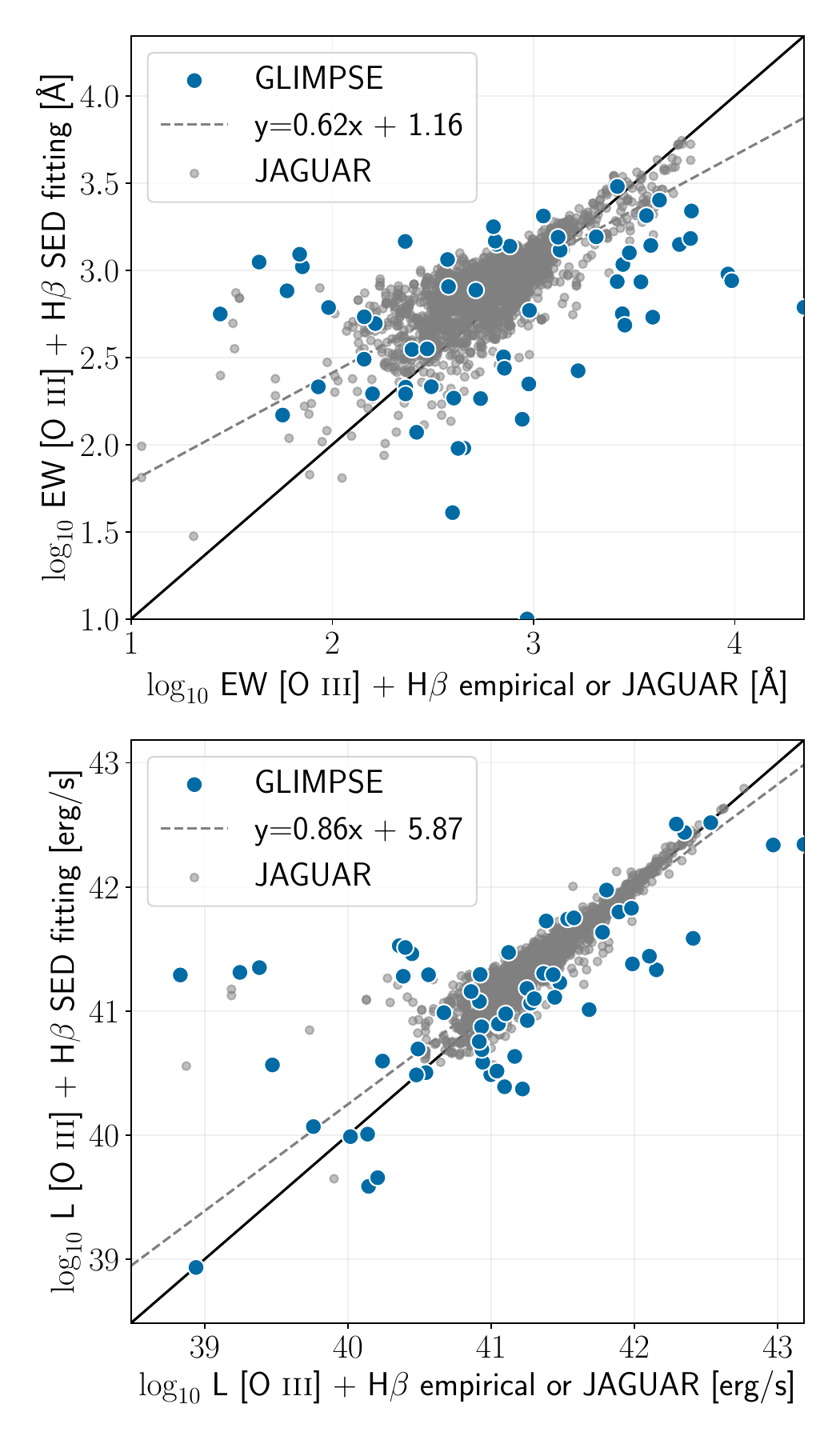}
    \caption{Comparison of measurement of the equivalent width and line flux of \oiiihb~using SED fitting (CIGALE) against the empirical method (blue dots) and the JAGUAR simulations (grey dots). Only sources with a valid empirical measurement were kept. For JAGUAR, we kept sources with redshift measurements within 1$\sigma$ to ensure a fairer comparison with GLIMPSE. We did not display uncertainties for readability reasons. For the JAGUAR simulation, we added a linear fit on the equivalent widths and line fluxes (dashed grey lines).}
    \label{fig:comp_cgl_emp}
\end{figure}

In Fig.~\ref{fig:comp_cgl_emp} we show the comparison for the measurement of the equivalent width and line flux of \oiiihb~using SED fitting for GLIMPSE and JAGUAR data. The GLIMPSE data compares against the empirical measurements described above, and the JAGUAR data against the truth values obtained by the simulation. While we observe some scatter in equivalent width measurements, the line fluxes match well for both methods. All the measurements correlate, except for the empirically reconstructed equivalent width. For the equivalent width, the correlation coefficients are 0.79 for JAGUAR and 0.20 for empirical, and for line fluxes, 0.93 for JAGUAR and 0.67 for empirical. The comparison to GLIMPSE observations is more scattered than the comparison to the simulation. This might come from the lower amount of information used by the empirical method, which reduces the constraints on the measurement. In general, we observe an average good agreement between CIGALE and the two methods, validating our use of CIGALE for the final measurements.

\section{Example of galaxies}
\label{app:sed_examples}
\begin{figure*}
    \centering
    \includegraphics[width=1\linewidth]{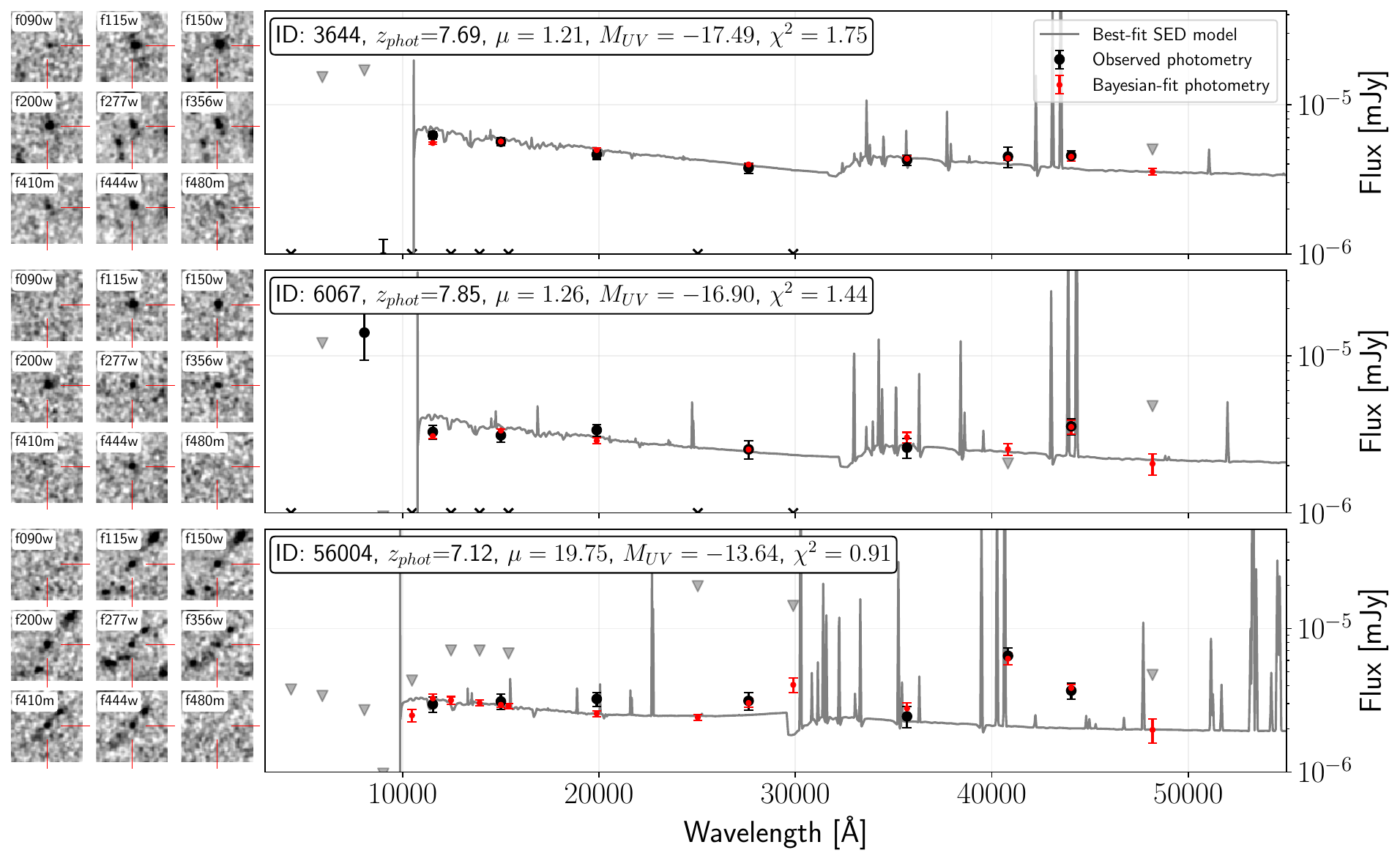}
    \caption{Galaxy stamp cutout and SED for three representative galaxies in the sample. Left: 3x3 panels show the cutouts for the nine GLIMPSE JWST filters. Right: the observed photometry for that galaxy (black), the photometry obtained using Bayesian inference of CIGALE (red) and the best fit SED (grey). The three galaxies were selected for being bright, faint without strong magnification, and very-faint with strong magnification.}
    \label{fig:sed_examples}
\end{figure*}

In Fig.\ \ref{fig:sed_examples} we display a few example SED fitting and galaxies cutouts. The three galaxies selected cover most situations: ID=3644 is a relatively bright galaxy with limited magnification, ID=6067 is a faint galaxy detected with limited magnification and finally ID=56004 is very-faint but is strongly magnified by the lensing field.

\section{Luminosity functions and associated data}
\label{app:LFs}
In this appendix we report the additional LFs discussed in the paper that did not fit in it, as well as the data of all the LFs in this paper. All the tables in this section include the measurement of $\log_{10}\Phi(L)$ for each luminosity bin $L$, for the two redshift bins. In addition, we report the average number of sources per bin, as well as  the average completeness of these sources, including uncertainties.

\subsection{\oiiihb~luminosity function}
\label{app:o3lf}
In Table~\ref{tab:O3LF} we report the \oiiihb~LF data from Fig.~\ref{fig:O3LF}. Only for this table, we report the average \Rthree~used to transform our \oiiihb~LF to \Oiiib~and \hb.\\

\begin{table*}[t]
    \centering
    \caption{LF measured in Fig.~\ref{fig:O3LF}.}
    \label{tab:O3LF}

    \begin{tabular}{lllllll}
\hline
 & $\log_{10}(L_{\oiiihb})$ & $\mean{N}$ & $\mean{C}$ & $\log_{10}\Phi(L)$ & S/N & $\mean{R3}$\\
  & erg / s &   &   & Mpc$^{-3}$dex$^{-1}$ &   &  \\
\hline
$\mean{z} \sim 7.48$ & 38.75 & $3.71 \pm 1.54$ & $0.14 \pm 0.18$ & $-0.33 \pm 0.69$ & $1.28$ & $2.69 \pm 0.42$\\
 & 39.25 & $8.40 \pm 2.32$ & $0.18 \pm 0.18$ & $-0.47 \pm 0.40$ & $2.41$ & $3.01 \pm 0.22$\\
 & 39.75 & $14.20 \pm 2.96$ & $0.23 \pm 0.16$ & $-0.64 \pm 0.33$ & $2.98$ & $3.29 \pm 0.13$\\
 & 40.25 & $22.99 \pm 3.53$ & $0.38 \pm 0.17$ & $-1.07 \pm 0.32$ & $3.05$ & $3.79 \pm 0.08$\\
 & 40.75 & $26.35 \pm 3.63$ & $0.47 \pm 0.15$ & $-1.62 \pm 0.22$ & $4.46$ & $4.24 \pm 0.19$\\
 & 41.25 & $27.51 \pm 3.07$ & $0.50 \pm 0.10$ & $-1.82 \pm 0.18$ & $5.37$ & $4.51 \pm 0.12$\\
 & 41.75 & $12.38 \pm 1.46$ & $0.61 \pm 0.07$ & $-2.37 \pm 0.26$ & $3.80$ & $5.98 \pm 0.17$\\
 & 42.25 & $4.34 \pm 0.93$ & $0.71 \pm 0.02$ & $-2.90 \pm 0.44$ & $2.15$ & $8.38 \pm 0.14$\\
 & 42.75 & $1.86 \pm 0.83$ & $0.72 \pm 0.01$ & $-3.28 \pm 0.71$ & $1.24$ & $8.50 \pm 0.00$\\
\hline
$\mean{z} \sim 8.33$ & 39.25 & $2.68 \pm 1.32$ & $0.25 \pm 0.13$ & $-1.66 \pm 1.00$ & $0.76$ & $3.42 \pm 0.34$\\
 & 39.75 & $4.69 \pm 1.63$ & $0.23 \pm 0.14$ & $-1.51 \pm 0.56$ & $1.66$ & $3.49 \pm 0.19$\\
 & 40.25 & $7.81 \pm 1.99$ & $0.30 \pm 0.12$ & $-1.66 \pm 0.53$ & $1.77$ & $3.78 \pm 0.11$\\
 & 40.75 & $5.84 \pm 1.93$ & $0.38 \pm 0.14$ & $-1.98 \pm 0.59$ & $1.56$ & $3.94 \pm 0.15$\\
 & 41.25 & $10.64 \pm 1.94$ & $0.46 \pm 0.12$ & $-2.21 \pm 0.30$ & $3.24$ & $4.71 \pm 0.23$\\
 & 41.75 & $8.10 \pm 1.35$ & $0.60 \pm 0.05$ & $-2.51 \pm 0.32$ & $3.01$ & $6.39 \pm 0.21$\\
\hline
\end{tabular}
\tablefoot{
$L_{\oiiihb}$ is the line luminosity, $\mean{N}$ is the number of sources in the luminosity bin, $\mean{C}$ is the mean completeness for the sources in the bin, $\Phi(L)$ is the LF, S/N is the signal to noise ratio and $\mean{R3}$ is the average \Rthree~ratio in the luminosity bin assuming Eq.~\ref{eq:varying_oiii_hb_ratio}.
}

\end{table*}

\subsection{\oiiihb~luminosity function with flux cut to $L\geq10^{41}$ \ergs}
\label{app:o3lf_woldlike}

\begin{figure*}[th]
   \centering
   \includegraphics[width=0.9\linewidth]{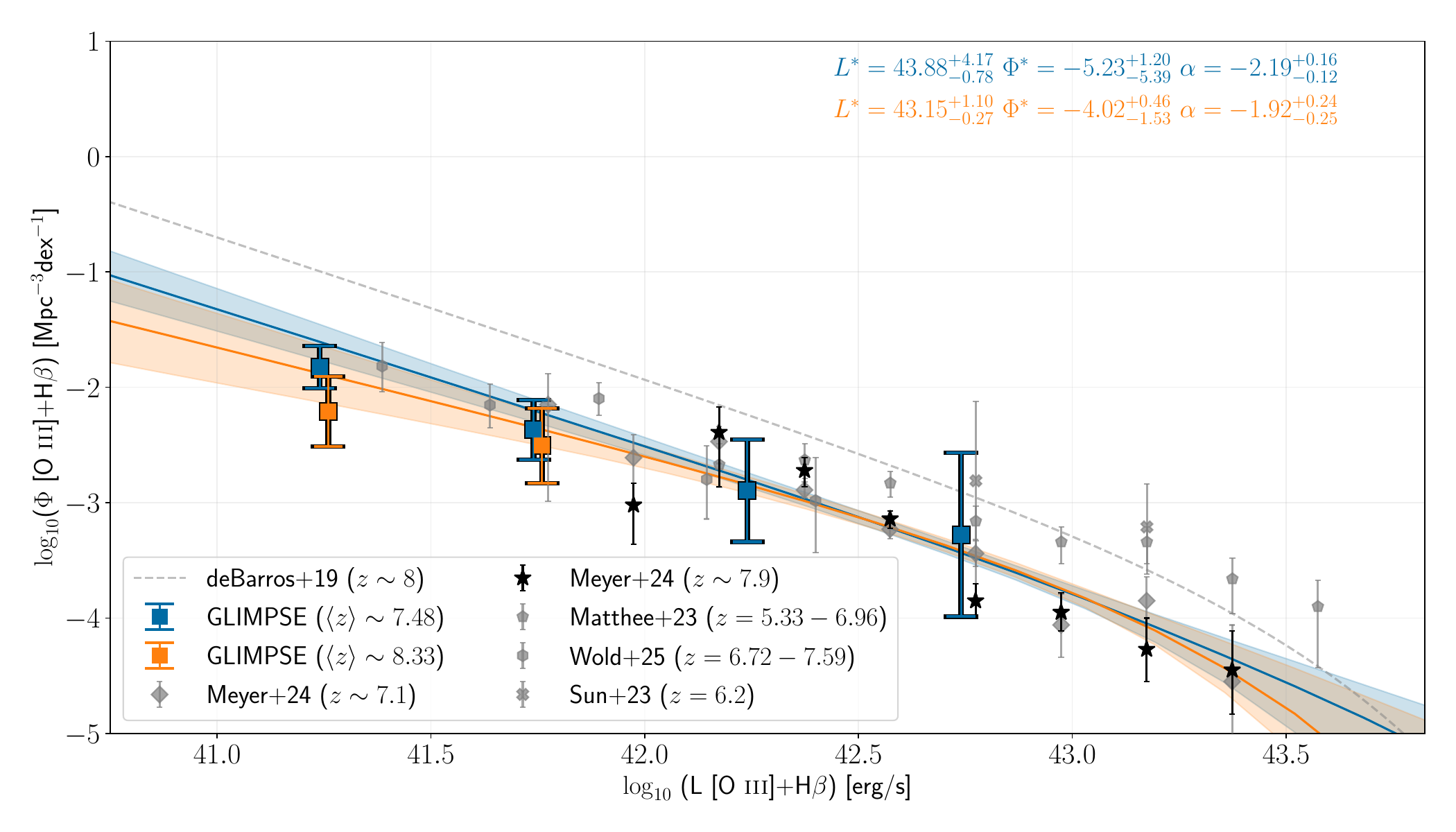}
    \caption{LF of \oiiihbfull~for the different redshift ranges considered. For the fit of the LF, we limited the flux bins to the depth of \citep{wold_uncovering_2025}, and discarded bins are displayed with a cross. We added previous studies of the LF using JWST/NIRCam GRISM instrument \citep{meyer_jwst_2024, matthee_eiger_2023}, a JWST/NIRCam wide field slitless spectroscopy study by \citep{sun_first_2023}, a JWST/NIRCam medium band survey by \citep{wold_uncovering_2025} and a former \textit{Spitzer} study by \cite{de_barros_greats_2019}. All the JWST surveys specifically study the \oiii$\lambda5007$Å, so we boosted their luminosity using their respective R3 ratio and $\Oiiib/\Oiiit=2.98$ \citep{storey_theoretical_2000}, according to our assumed line of ratios. This approximation matches the stacked median and flux measurements from these papers. The data can be found in Table~\ref{tab:O3LF_woldlike} and the parametrisation in Table~\ref{tab:schechter}.}
    \label{fig:O3LF_woldlike}
\end{figure*}

GLIMPSE probes \oiiihb\ emitters fainter than any previous study, making it difficult to compare directly.
Before GLIMPSE, \citet{wold_uncovering_2025} was the deepest study of the \oiiihb~LF (reaching $L_\oiiihb\sim10^{41}$\ergs). As their faint-end slope is much steeper than GLIMPSE ($\alpha = -2.07_{-0.23}^{+0.22}$), we truncate our \oiiihb~LF at $10^{41}$\ergs\ to compare both studies and examine the behaviour of the fainter galaxies. Figure~\ref{fig:O3LF_woldlike} shows the \oiiihb~LF fixed at the limited depth. We observe a similar faint-end slope, which shows that the fainter galaxy population flattens the LF. The associated data are provided in Table~\ref{tab:O3LF_woldlike}.\\

\begin{table}[t]
    \centering
    \tiny
    \caption{LF measured in Fig.~\ref{fig:O3LF_woldlike}.}
    \label{tab:O3LF_woldlike}

    \begin{tabular}{llllll}
\hline
 & $\log_{10}(L)$ & $\mean{N}$ & $\mean{C}$ & $\log_{10}\Phi(L)$ & S/N\\
  & erg / s &   &   & Mpc$^{-3}$dex$^{-1}$ &  \\
\hline
$\mean{z} \sim 7.48$ & 41.25 & $27.49 \pm 3.01$ & $0.50 \pm 0.10$ & $-1.82 \pm 0.18$ & $5.37$\\
 & 41.75 & $12.35 \pm 1.46$ & $0.61 \pm 0.07$ & $-2.37 \pm 0.26$ & $3.79$\\
 & 42.25 & $4.36 \pm 0.94$ & $0.71 \pm 0.02$ & $-2.90 \pm 0.44$ & $2.15$\\
 & 42.75 & $1.85 \pm 0.84$ & $0.72 \pm 0.01$ & $-3.28 \pm 0.71$ & $1.24$\\
\hline
$\mean{z} \sim 8.33$ & 41.25 & $10.61 \pm 1.97$ & $0.46 \pm 0.12$ & $-2.21 \pm 0.30$ & $3.23$\\
 & 41.75 & $8.12 \pm 1.34$ & $0.60 \pm 0.05$ & $-2.51 \pm 0.32$ & $3.01$\\
\hline
\end{tabular}
\tablefoot{
$L$ is the \oiiihb\ line luminosity, $\mean{N}$ is the number of sources in the luminosity bin, $\mean{C}$ is the mean completeness for the sources in the bin, $\Phi(L)$ is the LF and S/N is the signal to noise ratio.
}

\end{table}

\subsection{Separating \Oiiib~and \hb~from the \oiiihb~luminosity function}

\begin{figure*}
    \centering
    \includegraphics[width=0.9\linewidth]{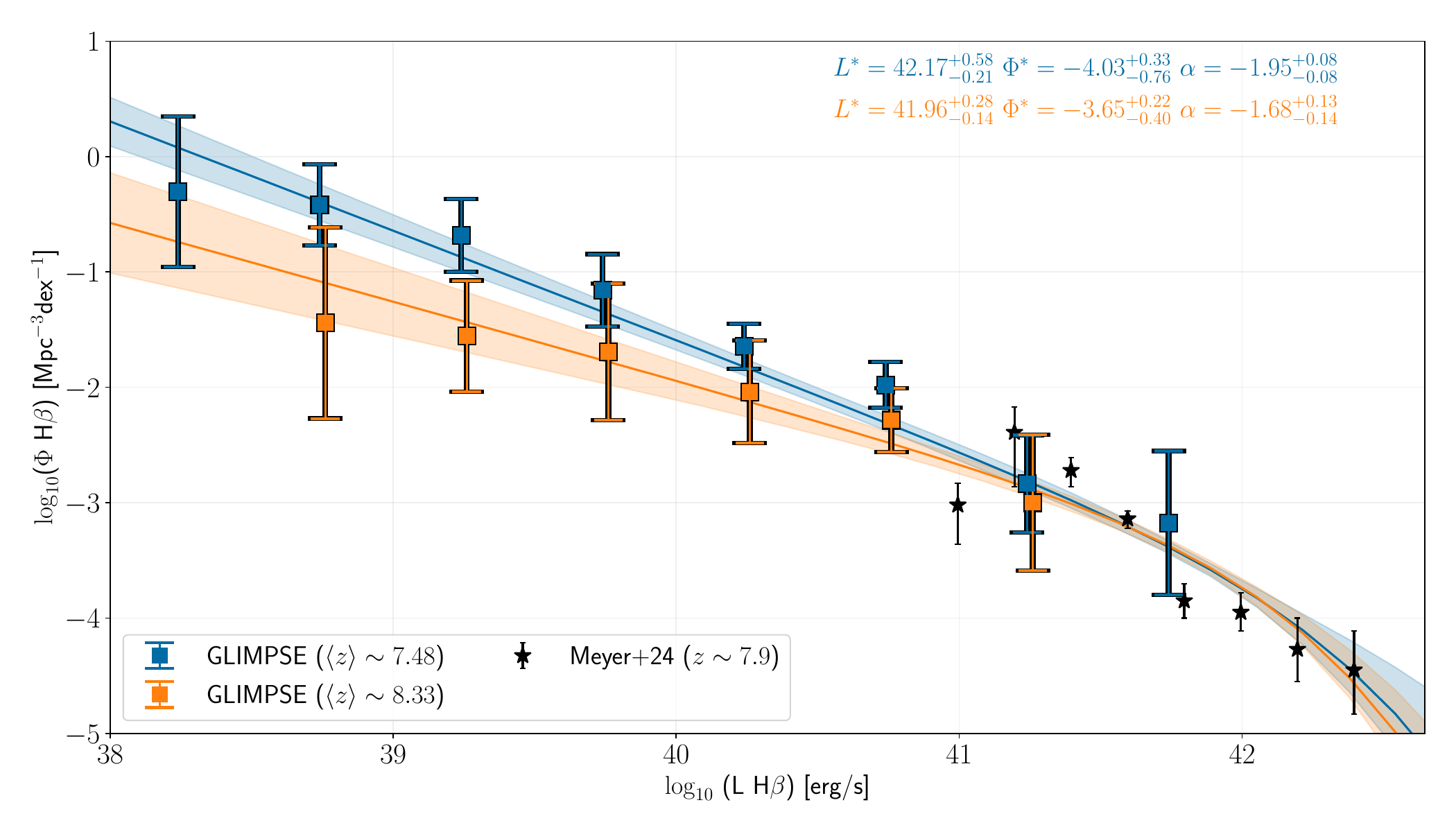}
    \caption{\hb~LF separated from the main \oiiihb~LF using the \Rthree ratio from Eq.~\ref{eq:varying_oiii_hb_ratio}. The \cite{meyer_jwst_2024} values are converted using their median \Rthree value. The data can be found in Table~\ref{tab:HbLF} and the parametrisation can be found in Table~\ref{tab:schechter}.}
    \label{fig:HbLF}
\end{figure*}

\begin{figure*}
    \centering
    \includegraphics[width=0.9\linewidth]{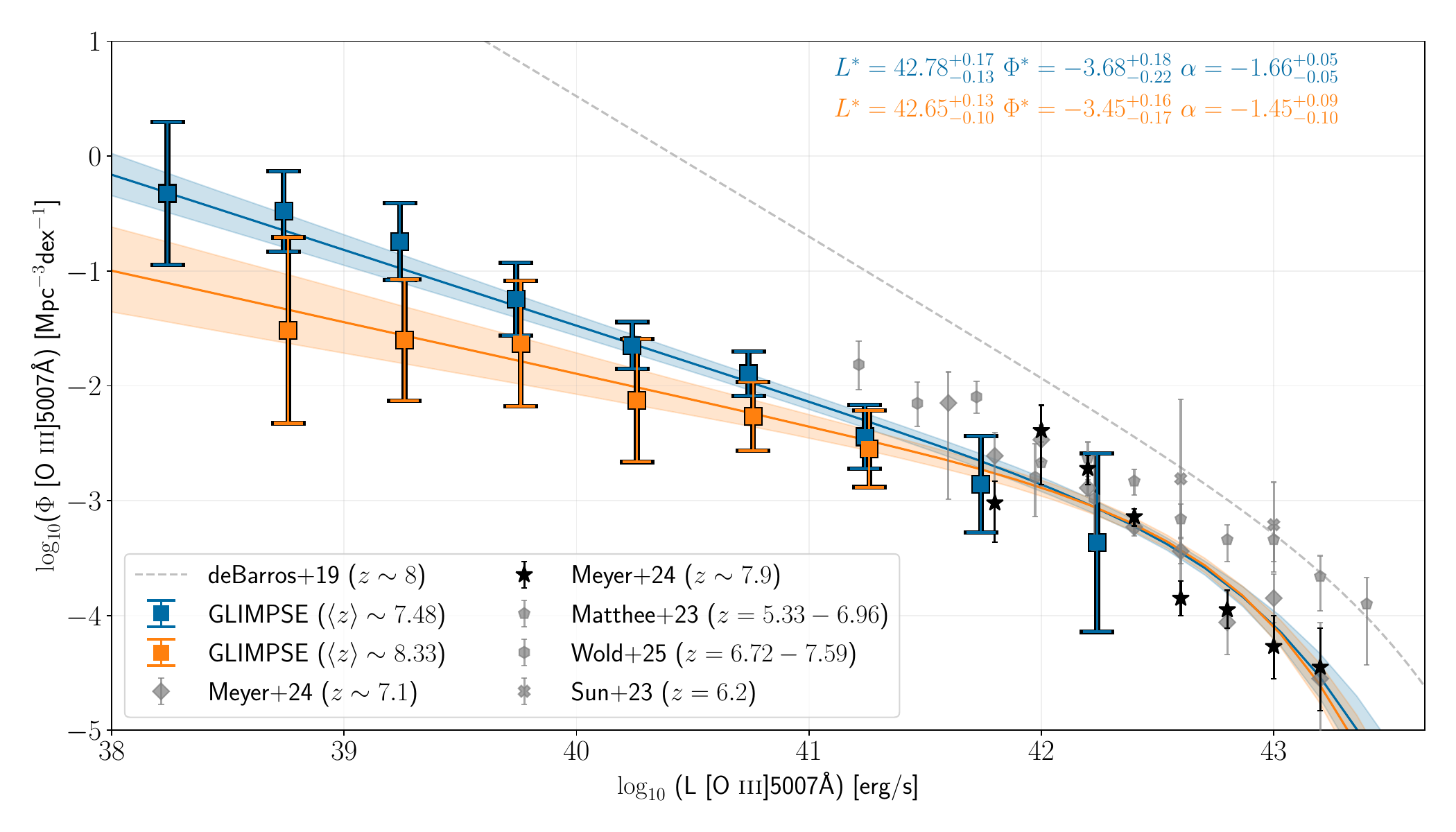}
    \caption{\Oiiib~LF separated from the main \oiiihb~LF using the \Rthree ratio from Eq.~\ref{eq:varying_oiii_hb_ratio}. The data can be found in Table~\ref{tab:O35007LF} and the parametrisation can be found in Table~\ref{tab:schechter}.}
    \label{fig:O35007LF}
\end{figure*}

Differences in metallicity varies the \Rthree~ratio \citep[e.g.][]{maiolino_re_2019}, which can explain differences between the \oiiihb~LF and the UV LF. By assuming an evolving R3 ratio following Eq.~\ref{eq:varying_oiii_hb_ratio}, we can separate the contribution of \hb~and \oiii in the \oiiihb~LF (see Sect.~\ref{sec:impact_Z} for more detail and Table~\ref{tab:O3LF} for the average \Rthree~measurement in each luminosity bin). Figures~\ref{fig:HbLF} and~\ref{fig:O35007LF} show the LF for \hb~and \Oiiib. Their associated data can be found in Tables~\ref{tab:HbLF} and~\ref{tab:O35007LF}.\\

\begin{table}[t]
    \centering
    \tiny
    \caption{\hb~luminosity function from Fig.~\ref{fig:HbLF}.}
    \label{tab:HbLF}

    \begin{tabular}{llllll}
\hline
 & $\log_{10}(L)$ & $\mean{N}$ & $\mean{C}$ & $\log_{10}\Phi(L)$ & S/N\\
  & erg / s &   &   & Mpc$^{-3}$dex$^{-1}$ &  \\
\hline
$\mean{z} \sim 7.48$ & 38.25 & $4.15 \pm 1.62$ & $0.16 \pm 0.21$ & $-0.30 \pm 0.65$ & $1.38$\\
 & 38.75 & $10.77 \pm 2.55$ & $0.20 \pm 0.18$ & $-0.42 \pm 0.35$ & $2.75$\\
 & 39.25 & $17.93 \pm 3.30$ & $0.27 \pm 0.17$ & $-0.68 \pm 0.32$ & $3.09$\\
 & 39.75 & $28.29 \pm 3.63$ & $0.44 \pm 0.17$ & $-1.16 \pm 0.31$ & $3.11$\\
 & 40.25 & $29.64 \pm 3.49$ & $0.47 \pm 0.13$ & $-1.64 \pm 0.20$ & $5.08$\\
 & 40.75 & $23.19 \pm 2.33$ & $0.55 \pm 0.11$ & $-1.98 \pm 0.20$ & $5.01$\\
 & 41.25 & $4.91 \pm 1.14$ & $0.69 \pm 0.04$ & $-2.84 \pm 0.42$ & $2.26$\\
 & 41.75 & $2.38 \pm 0.75$ & $0.72 \pm 0.02$ & $-3.18 \pm 0.62$ & $1.45$\\
\hline
$\mean{z} \sim 8.33$ & 38.75 & $3.79 \pm 1.50$ & $0.26 \pm 0.14$ & $-1.44 \pm 0.83$ & $1.01$\\
 & 39.25 & $6.43 \pm 1.83$ & $0.24 \pm 0.13$ & $-1.56 \pm 0.48$ & $1.96$\\
 & 39.75 & $7.28 \pm 2.00$ & $0.35 \pm 0.14$ & $-1.69 \pm 0.59$ & $1.54$\\
 & 40.25 & $8.25 \pm 1.99$ & $0.40 \pm 0.13$ & $-2.04 \pm 0.45$ & $2.13$\\
 & 40.75 & $11.61 \pm 1.75$ & $0.55 \pm 0.11$ & $-2.28 \pm 0.28$ & $3.57$\\
 & 41.25 & $2.75 \pm 0.90$ & $0.61 \pm 0.02$ & $-3.00 \pm 0.59$ & $1.56$\\
\hline
\end{tabular}
\tablefoot{$L$ is the \hb\ line luminosity, $\mean{N}$ is the number of sources in the luminosity bin, $\mean{C}$ is the mean completeness for the sources in the bin, $\Phi(L)$ is the LF and S/N is the signal to noise ratio.}

\end{table}

\begin{table}[t]
    \centering
    \tiny
    \caption{\Oiiib~LF from Fig.~\ref{fig:O35007LF}.}
    \label{tab:O35007LF}
    
    \begin{tabular}{llllll}
\hline
 & $\log_{10}(L)$ & $\mean{N}$ & $\mean{C}$ & $\log_{10}\Phi(L)$ & S/N\\
  & erg / s &   &   & Mpc$^{-3}$dex$^{-1}$ &  \\
\hline
$\mean{z} \sim 7.48$ & 38.25 & $4.33 \pm 1.69$ & $0.13 \pm 0.17$ & $-0.33 \pm 0.62$ & $1.45$\\
 & 38.75 & $9.87 \pm 2.42$ & $0.19 \pm 0.18$ & $-0.48 \pm 0.35$ & $2.76$\\
 & 39.25 & $14.66 \pm 3.04$ & $0.25 \pm 0.15$ & $-0.75 \pm 0.34$ & $2.90$\\
 & 39.75 & $23.37 \pm 3.57$ & $0.41 \pm 0.16$ & $-1.24 \pm 0.32$ & $3.07$\\
 & 40.25 & $26.98 \pm 3.47$ & $0.47 \pm 0.15$ & $-1.65 \pm 0.20$ & $4.85$\\
 & 40.75 & $25.24 \pm 2.89$ & $0.51 \pm 0.10$ & $-1.90 \pm 0.19$ & $5.16$\\
 & 41.25 & $10.68 \pm 1.34$ & $0.63 \pm 0.06$ & $-2.44 \pm 0.28$ & $3.53$\\
 & 41.75 & $4.77 \pm 0.90$ & $0.71 \pm 0.02$ & $-2.86 \pm 0.42$ & $2.28$\\
 & 42.25 & $1.34 \pm 0.85$ & $0.72 \pm 0.01$ & $-3.37 \pm 0.78$ & $1.10$\\
\hline
$\mean{z} \sim 8.33$ & 38.75 & $3.25 \pm 1.35$ & $0.24 \pm 0.13$ & $-1.52 \pm 0.81$ & $1.04$\\
 & 39.25 & $5.19 \pm 1.73$ & $0.24 \pm 0.13$ & $-1.60 \pm 0.53$ & $1.77$\\
 & 39.75 & $7.57 \pm 2.01$ & $0.31 \pm 0.13$ & $-1.63 \pm 0.55$ & $1.70$\\
 & 40.25 & $6.13 \pm 1.94$ & $0.38 \pm 0.13$ & $-2.13 \pm 0.53$ & $1.74$\\
 & 40.75 & $10.37 \pm 1.86$ & $0.48 \pm 0.11$ & $-2.27 \pm 0.30$ & $3.27$\\
 & 41.25 & $7.41 \pm 1.13$ & $0.61 \pm 0.04$ & $-2.55 \pm 0.33$ & $2.90$\\
\hline
\end{tabular}
\tablefoot{$L$ is the \Oiiib\ line luminosity, $\mean{N}$ is the number of sources in the luminosity bin, $\mean{C}$ is the mean completeness for the sources in the bin, $\Phi(L)$ is the LF and S/N is the signal to noise ratio.}

\end{table}

\subsection{Dust correction to obtain the \ha~luminosity function}
\begin{figure*}[htb]
    \centering
    \includegraphics[width=0.9\linewidth]{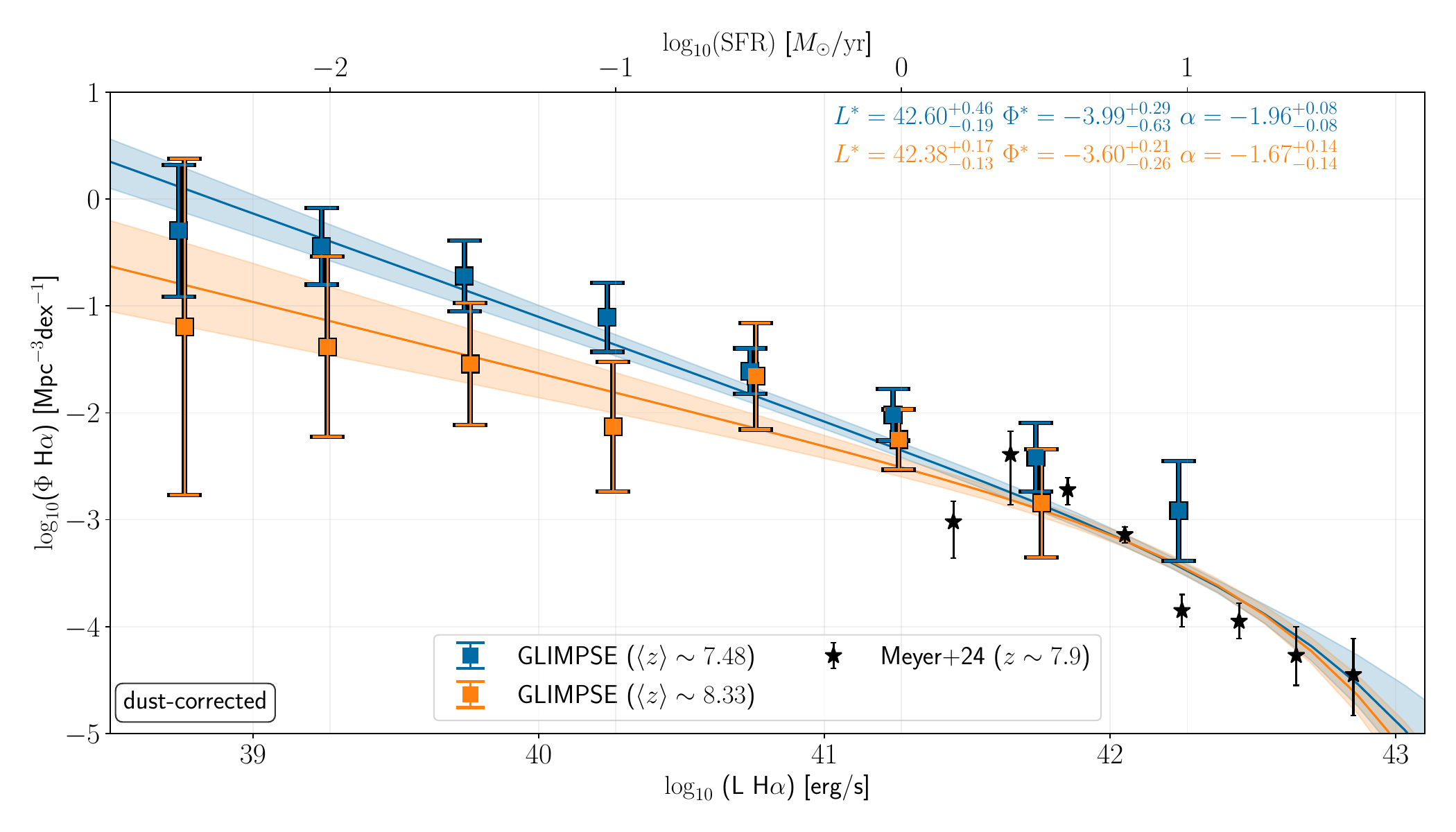}
    \caption{Dust-corrected \ha~LF obtained by converting \oiiihb~using the \Rthree ratio given by Eq.~\ref{eq:varying_oiii_hb_ratio}. The data from \citep{meyer_jwst_2024} was converted using their median \Rthree, assuming no dust attenuation as it was measured as negligible. Both used the usual $\ha/\hb = 2.86$ \citep{osterbrock_astrophysics_2006}. We added the \citep{kennicutt_star_2012} conversion of \ha~to SFR on top for a better understanding of the reached fluxes and implications. The data can be found in Table~\ref{tab:HaLF} and the parametrisation in Table~\ref{tab:schechter}.}
    \label{fig:HaLF}
\end{figure*}

To measure the ionising photon-production rate and the CSFRD, we need to correct the separated \hb~LF (Fig.~\ref{fig:HbLF}) for dust attenuation. Indeed, \oiii~and \hb~are very close, and therefore, the dust attenuation is similar. However, \ha~is affected differently from \hb, which requires a correction. The detail of the correction can be found in Sect.~\ref{sec:impl_ion}. Figure~\ref{fig:HaLF} shows the dust-corrected \ha~LF with its associated data in Table~\ref{tab:HaLF}.

\begin{table}[t]
    \centering
    \tiny
    \caption{Dust-corrected \ha~LF from Fig.~\ref{fig:HaLF}.}

\begin{tabular}{llllll}
\hline
 & $\log_{10}(L)$ & $\mean{N}$ & $\mean{C}$ & $\log_{10}\Phi(L)$ & S/N\\
  & erg / s &   &   & Mpc$^{-3}$dex$^{-1}$ &  \\
\hline
$\mean{z} \sim 7.48$ & 38.75 & $4.35 \pm 1.65$ & $0.16 \pm 0.22$ & $-0.30 \pm 0.62$ & $1.47$\\
 & 39.25 & $10.49 \pm 2.52$ & $0.20 \pm 0.18$ & $-0.44 \pm 0.36$ & $2.70$\\
 & 39.75 & $16.80 \pm 3.24$ & $0.28 \pm 0.18$ & $-0.72 \pm 0.33$ & $2.92$\\
 & 40.25 & $27.08 \pm 3.63$ & $0.43 \pm 0.17$ & $-1.11 \pm 0.32$ & $3.02$\\
 & 40.75 & $30.06 \pm 3.60$ & $0.46 \pm 0.13$ & $-1.61 \pm 0.21$ & $4.68$\\
 & 41.25 & $18.83 \pm 2.60$ & $0.54 \pm 0.12$ & $-2.02 \pm 0.24$ & $4.09$\\
 & 41.75 & $9.47 \pm 1.56$ & $0.60 \pm 0.12$ & $-2.41 \pm 0.32$ & $3.03$\\
 & 42.25 & $4.07 \pm 0.92$ & $0.70 \pm 0.05$ & $-2.92 \pm 0.47$ & $2.03$\\
\hline
$\mean{z} \sim 8.33$ & 38.75 & $1.11 \pm 0.98$ & $0.19 \pm 0.19$ & $-1.20 \pm 1.57$ & $0.30$\\
 & 39.25 & $3.68 \pm 1.43$ & $0.24 \pm 0.14$ & $-1.38 \pm 0.84$ & $0.98$\\
 & 39.75 & $6.69 \pm 1.84$ & $0.25 \pm 0.13$ & $-1.54 \pm 0.57$ & $1.61$\\
 & 40.25 & $5.73 \pm 1.89$ & $0.37 \pm 0.13$ & $-2.13 \pm 0.61$ & $1.50$\\
 & 40.75 & $8.46 \pm 1.98$ & $0.38 \pm 0.14$ & $-1.66 \pm 0.50$ & $1.88$\\
 & 41.25 & $11.83 \pm 1.74$ & $0.53 \pm 0.11$ & $-2.25 \pm 0.28$ & $3.48$\\
 & 41.75 & $3.63 \pm 0.98$ & $0.59 \pm 0.06$ & $-2.85 \pm 0.50$ & $1.86$\\
\hline
\end{tabular}

\tablefoot{$L$ is the \ha\ line luminosity, $\mean{N}$ is the number of sources in the luminosity bin, $\mean{C}$ is the mean completeness for the sources in the bin, $\Phi(L)$ is the LF and S/N is the signal to noise ratio.}
    
    \label{tab:HaLF}
\end{table}

\section{Faint-end slope compilation}
We report in Table~\ref{tab:alpha} the non-exhaustive list of measurement of the faint-end slope $\alpha$ for several nebular tracers, as well as UV. The measurement of GLIMPSE are already reported in Table~\ref{tab:schechter}.

\begin{table}[]
    \centering
    \tiny
    \caption{Non-exhaustive compilation of $\alpha$ measurements for \oiii, \hb, \ha~and UV LFs.}
    \label{tab:alpha}
\begin{tabular}{@{}lllll@{}}
\hline
Paper & Redshift & Tracer & Par. & $\alpha$\\
\hline
\citet{sobral_large_2013} & 0.40 & H$\alpha$ & SCH & $-1.75_{-0.08}^{+0.12}$\\
  & 0.84 & H$\alpha$ & SCH & $-1.56_{-0.14}^{+0.13}$\\
  & 1.47 & H$\alpha$ & SCH & $-1.62_{-0.29}^{+0.25}$\\
  & 2.23 & H$\alpha$ & SCH & $-1.59_{-0.13}^{+0.12}$\\
\citet{colbert_predicting_2013} & 0.60 & H$\alpha$ & SCH & $-1.27_{-0.12}^{+0.17}$\\
  & 1.20 & H$\alpha$ & SCH & $-1.43_{-0.12}^{+0.17}$\\
\citet{hayashi_16_2018} & 0.25 & H$\alpha$ & SCH & $-1.59_{-0.05}^{+0.05}$\\
  & 0.40 & H$\alpha$ & SCH & $-1.75_{-0.06}^{+0.06}$\\
\citet{khostovan_large_2020} & 0.47 & H$\alpha$ & SCH & $-1.77_{-0.11}^{+0.12}$\\
\citet{nagaraj_h_2023} & 1.36 & H$\alpha$ + [N {\sc ii}] & SCH & $-1.60_{-0.07}^{+0.07}$\\
\citet{bollo_h_2023} & 4.50 & H$\alpha$ & SCH & $-1.83_{-0.09}^{+0.07}$\\
\citet{covelo-paz_h_2025} & 4.45 & H$\alpha$ & SCH & $-1.64_{-0.21}^{+0.27}$\\
  & 5.30 & H$\alpha$ & SCH & $-1.58_{-0.25}^{+0.28}$\\
  & 6.15 & H$\alpha$ & SCH & $-1.49_{-0.33}^{+0.36}$\\
\citet{fu_medium-band_2025} & 4.50 & H$\alpha$ & SCH & $-1.83_{-0.13}^{+0.13}$\\
  & 6.30 & H$\alpha$ & SCH & $-1.85_{-0.19}^{+0.33}$\\
\citet{colbert_predicting_2013} & 1.10 & [O {\sc iii}] & SCH & $-1.40_{-0.15}^{+0.78}$\\
  & 1.90 & [O {\sc iii}] & SCH & $-1.67_{-0.15}^{+0.78}$\\
\citet{hayashi_16_2018} & 0.63 & [O {\sc iii}] & SCH & $-1.42_{-0.14}^{+0.14}$\\
  & 0.84 & [O {\sc iii}] & SCH & $-1.95_{-0.11}^{+0.11}$\\
\citet{khostovan_large_2020} & 0.93 & [O {\sc iii}] & SCH & $-1.57_{-0.30}^{+0.35}$\\
\citet{bongiovanni_otelo_2020} & 0.83 & [O {\sc iii}] & SCH & $-1.03_{-0.08}^{+0.08}$\\
\citet{bowman_z_2021} & 2.12 & [O {\sc iii}] & SCH & $-1.51_{-0.28}^{+0.77}$\\
\citet{nagaraj_h_2023} & 1.53 & [O {\sc iii}]$\lambda$5008 & SCH & $-1.50_{-0.07}^{+0.07}$\\
\citet{wold_uncovering_2025} & 7.00 & [O {\sc iii}]$\lambda$5008 & DPL & $-2.07_{-0.23}^{+0.22}$\\
\citet{reddy_steep_2009} & 2.30 & UV & SCH & $-1.73_{-0.07}^{+0.07}$\\
  & 3.05 & UV & SCH & $-1.73_{-0.13}^{+0.13}$\\
\citet{parsa_galaxy_2016} & 1.70 & UV & SCH & $-1.33_{-0.03}^{+0.03}$\\
  & 1.90 & UV & SCH & $-1.32_{-0.03}^{+0.03}$\\
  & 2.25 & UV & SCH & $-1.26_{-0.04}^{+0.04}$\\
  & 2.80 & UV & SCH & $-1.31_{-0.04}^{+0.04}$\\
  & 3.80 & UV & SCH & $-1.43_{-0.04}^{+0.04}$\\
\citet{atek_extreme_2018} & 6.00 & UV & SCH & $-1.94_{-0.09}^{+0.11}$\\
\citet{moutard_uv_2020} & 0.17 & UV & SCH & $-1.41_{-0.02}^{+0.02}$\\
  & 0.38 & UV & SCH & $-1.37_{-0.02}^{+0.02}$\\
  & 0.53 & UV & SCH & $-1.41_{-0.05}^{+0.05}$\\
  & 0.75 & UV & SCH & $-1.40_{-0.04}^{+0.04}$\\
  & 1.10 & UV & SCH & $-1.43_{-0.07}^{+0.07}$\\
  & 1.55 & UV & SCH & $-1.45_{-0.07}^{+0.07}$\\
\citet{bowler_lack_2020} & 8.00 & UV & SCH & $-2.18_{-0.16}^{+0.16}$\\
  & 9.00 & UV & SCH & $-2.31_{-0.24}^{+0.24}$\\
\citet{bowler_lack_2020} & 8.00 & UV & DPL & $-1.96_{-0.15}^{+0.15}$\\
\citet{bouwens_z_2022} & 2.00 & UV & SCH & $-1.53_{-0.03}^{+0.03}$\\
  & 3.00 & UV & SCH & $-1.60_{-0.03}^{+0.03}$\\
  & 4.00 & UV & SCH & $-1.69_{-0.03}^{+0.03}$\\
  & 5.00 & UV & SCH & $-1.78_{-0.04}^{+0.04}$\\
  & 6.00 & UV & SCH & $-1.87_{-0.04}^{+0.04}$\\
  & 7.00 & UV & SCH & $-2.05_{-0.06}^{+0.06}$\\
  & 8.00 & UV & SCH & $-2.20_{-0.09}^{+0.09}$\\
  & 9.00 & UV & SCH & $-2.28_{-0.10}^{+0.10}$\\
\citet{donnan_evolution_2023} & 8.00 & UV & DPL & $-2.04_{-0.29}^{+0.29}$\\
\citet{harikane_pure_2024} & 7.00 & UV & SCH & $-1.97_{-0.12}^{+0.14}$\\
  & 8.00 & UV & SCH & $-2.16_{-0.21}^{+0.24}$\\

\hline
\end{tabular}
\end{table}

\begin{table}[]
    \centering
    \tiny
\begin{tabular}{@{}lllll@{}}
\hline
Paper & Redshift & Tracer & Par. & $\alpha$\\
\hline
\citet{harikane_pure_2024} & 7.00 & UV & DPL & $-2.08_{-0.11}^{+0.12}$\\
  & 8.00 & UV & DPL & $-2.27_{-0.25}^{+0.16}$\\
\citet{donnan_jwst_2024} & 9.00 & UV & DPL & $-2.00_{-0.47}^{+0.47}$\\
  & 10.00 & UV & DPL & $-1.98_{-0.40}^{+0.40}$\\
  & 11.00 & UV & DPL & $-2.19_{-0.69}^{+0.69}$\\
\citet{sun_ultraviolet_2024} & 0.70 & UV & SCH & $-1.32_{-0.06}^{+0.06}$\\
  & 0.90 & UV & SCH & $-1.42_{-0.07}^{+0.07}$\\
\citet{willott_steep_2024} & 8.00 & UV & SCH & $-2.04_{-0.24}^{+0.30}$\\
\citet{finkelstein_complete_2024} & 9.00 & UV & DPL & $-2.20_{-0.30}^{+0.40}$\\
  & 11.00 & UV & DPL & $-2.20_{-0.40}^{+0.60}$\\
  & 14.00 & UV & DPL & $-2.55_{-1.40}^{+1.05}$\\
\citet{chemerynska_first_2026} & 9.50 & UV & DPL & $-2.00_{-0.09}^{+0.09}$\\
  & 10.50 & UV & DPL & $-2.14_{-0.08}^{+0.06}$\\
  & 11.50 & UV & DPL & $-2.17_{-0.10}^{+0.13}$\\
  & 13.00 & UV & DPL & $-2.06_{-0.12}^{+0.08}$\\
\citet{weibel_exploring_2025} & 10.00 & UV & DPL & $-2.98_{-0.60}^{+0.84}$\\
\hline
\end{tabular}
\tablefoot{Two parametrisations are considered: the most common is the Schechter function (SCH), but we also report the use of double power law (DPL) in a few studies. The data is shown in Fig.~\ref{fig:alpha_evo}.}
\end{table}

\section{Cosmic star formation rate density}
We report in Table~\ref{tab:SFRD} the non-exhaustive list of measurement of SFRD from the literature, as well as the associated measurement from GLIMPSE. We limited the compilation to a few papers with lower integration limit nearby 0.3\msunyr, as the average evolution of SFRD is well known. Every measurement was converted to the \citet{chabrier_galactic_2003} IMF. The data is shown in Fig.~\ref{fig:SFRD}.

\begin{table}[t]
    \centering
    \tiny
    \caption{Non-exhaustive list of SFRD measurements shown in Fig.~\ref{fig:SFRD} from UV, \ha~and \oiiihb~between $z\sim0-10$.}
    \label{tab:SFRD}

    \begin{tabular}{lllll}
\hline
Paper & Redshift & SFRD & Tracer & Int. lim.\\
 &  & M$_\odot$/yr/Mpc$^3$ &  & M$_\odot$/yr\\
\hline
\citet{oesch_dearth_2018} & $3.80$ & $-1.11_{-0.12}^{+0.13}$ & UV & $0.30$\\
 & $4.90$ & $-1.39_{-0.11}^{+0.12}$ & UV & $0.30$\\
 & $5.90$ & $-1.64_{-0.12}^{+0.14}$ & UV & $0.30$\\
 & $6.80$ & $-1.88_{-0.07}^{+0.07}$ & UV & $0.30$\\
 & $7.90$ & $-2.20_{-0.07}^{+0.06}$ & UV & $0.30$\\
 & $10.20$ & $-3.28_{-0.16}^{+0.16}$ & UV & $0.30$\\
\citet{bouwens_alma_2020} & $3.00$ & $-1.16_{-0.09}^{+0.09}$ & UV & $0.30$\\
 & $3.80$ & $-1.24_{-0.06}^{+0.06}$ & UV & $0.30$\\
 & $4.90$ & $-1.53_{-0.06}^{+0.06}$ & UV & $0.30$\\
 & $5.90$ & $-1.85_{-0.06}^{+0.06}$ & UV & $0.30$\\
 & $6.80$ & $-2.10_{-0.06}^{+0.06}$ & UV & $0.30$\\
 & $7.90$ & $-2.42_{-0.06}^{+0.06}$ & UV & $0.30$\\
 & $10.40$ & $-3.28_{-0.45}^{+0.36}$ & UV & $0.30$\\
\citet{bollo_h_2023} & $4.50$ & $-1.26_{-0.06}^{+0.23}$ & H$\alpha$ & $0.27$\\
\citet{covelo-paz_h_2025} & $4.45$ & $-1.38_{-0.07}^{+0.05}$ & H$\alpha$ & $0.27$\\
 & $5.30$ & $-1.54_{-0.06}^{+0.04}$ & H$\alpha$ & $0.27$\\
 & $6.15$ & $-1.92_{-0.12}^{+0.10}$ & H$\alpha$ & $0.27$\\
\citet{fu_medium-band_2025} & $4.50$ & $-1.23_{-0.05}^{+0.05}$ & H$\alpha$ & $0.24$\\
 & $6.30$ & $-1.60_{-0.14}^{+0.13}$ & H$\alpha$ & $0.24$\\
\citet{khostovan_evolution_2015} & $0.84$ & $-1.26_{-0.04}^{+0.02}$ & [O {\sc iii}]+H$\beta$ & \textemdash\\
 & $1.42$ & $-1.08_{-0.06}^{+0.06}$ & [O {\sc iii}]+H$\beta$ & \textemdash\\
 & $2.23$ & $-0.97_{-0.12}^{+0.12}$ & [O {\sc iii}]+H$\beta$ & \textemdash\\
 & $3.24$ & $-1.07_{-0.11}^{+0.11}$ & [O {\sc iii}]+H$\beta$ & \textemdash\\
\hline
GLIMPSE (this work) & $7.48 \pm 0.29$ & $-2.12_{-0.05}^{+0.06}$ & [O {\sc iii}]+H$\beta$ & 0.3\\
 & $8.33 \pm 0.26$ & $-2.22_{-0.05}^{+0.06}$ & [O {\sc iii}]+H$\beta$ & 0.3\\
 & $7.48 \pm 0.29$ & $-1.84_{-0.07}^{+0.08}$ & [O {\sc iii}]+H$\beta$ & $0.005$\\
 & $8.33 \pm 0.26$ & $-2.09_{-0.08}^{+0.09}$ & [O {\sc iii}]+H$\beta$ & $0.005$\\
\hline
\end{tabular}
\tablefoot{The measurement from this work are given at the end of the list. All values are IMF-corrected to \citet{chabrier_galactic_2003}. GLIMPSE measurements include both SFR integration limits: the standard 0.3\msunyr and the deeper GLIMPSE 0.005\msunyr ($L_{\ha}\sim 10^{39}$\ergs).}

\end{table}

\end{document}